\documentclass[review]{elsarticle}
\pdfoutput=1
\usepackage{booktabs} 
\usepackage{natbib}
\usepackage[ruled]{algorithm2e} 

\usepackage[disable]{todonotes}
\usepackage{longtable}
\usepackage{url}
\usepackage{tabularx}
\usepackage{graphicx}
\usepackage{adjustbox}
\usepackage{rotating}
\usepackage{pdfpages}

\newcommand{\mael}[1]{\todo[color=red!40, inline]{\footnotesize{Mael: #1}}}

\usepackage{amssymb}

\begin{document}

\newcommand{\RQOne}{What definitions of time pressure are used?}
\newcommand{\RQTwo}{What metrics are used to measure time pressure?}
\newcommand{\RQThree}{What process phases or approaches are studied with respect to time pressure?}
\newcommand{\RQFour}{What causes of time pressure are reported?}
\newcommand{\RQFive}{What are the effects of time pressure on software development?}

\title{Time Pressure in Software Engineering: A Systematic Review}

\author[1]{Miikka Kuutila\corref{cor1}}
\ead{miikka.kuutila@oulu.fi}
\author[1]{Mika M{\"a}ntyl{\"a}}
\author[1]{Umar Farooq}
\author[1]{Ma{\"e}lick Claes}
\address[1]{University of Oulu, M3S, Pentti Kaiteran katu 1, 90014 Oulu, Finland}
\cortext[cor1]{Corresponding author}

\begin{abstract}
\textbf{Context}: Large project overruns and overtime work have been reported in the software industry, resulting in additional expense for companies and personal issues for developers. Experiments and case studies have investigated the relationship between time pressure and software quality and productivity.
\textbf{Objective}: The present work aims to provide an overview of studies related to time pressure in software engineering; specifically, existing definitions, possible causes, and metrics relevant to time pressure were collected, and a mapping of the studies to software processes and approaches was performed. Moreover, we synthesize results of existing quantitative studies on the effects of time pressure on software development, and offer practical takeaways for practitioners and researchers, based on empirical evidence.
\textbf{Method}: Our search strategy examined 5,414 sources, found through repository searches and snowballing. Applying inclusion and exclusion criteria resulted in the selection of 102 papers, which made relevant contributions related to time pressure in software engineering.
\textbf{Results}: The majority of high quality studies report increased productivity and decreased quality under time pressure. The most frequent categories of studies focus on quality assurance, cost estimation, and process simulation. It appears that time pressure is usually caused by errors in cost estimation. The effect of time pressure is most often identified during software quality assurance. 
\textbf{Conclusions}: The majority of empirical studies report increased productivity under time pressure, while the most cost estimation and process simulation models assume that compressing the schedule increases the total needed hours. We also find evidence of the mediating effect of knowledge on the effects of time pressure, and that tight deadlines impact tasks with an algorithmic nature more severely. Future research should better contextualize quantitative studies to account for the existing conflicting results and to provide an understanding of situations when time pressure is either beneficial or harmful.
\end{abstract}

\maketitle

\section{Introduction}\label{sec:introduction}
Interest in scheduling and time-related issues in software engineering has been expressed for decades. In the 1970s, in the widely influential \emph{The Mythical Man-Month: Essays on Software Engineering}, Frederick Brooks coined the idea known as Brooks' law: ``adding manpower to a late software project makes it later"~\cite{brooks1995mythical}. Similarly, textbooks from the '80s and '90s for software developers and managers, have dedicated chapters and subchapters for ``deadline pressure" and ``beating schedule pressure"~\cite{gilb1988principles, mcconnell1996rapid}. More recently, it has been shown that 60-80\% of software projects are late (encounter overruns)~\cite{molokken2003review}; because being late is an antecedent of time pressure, we can assume the latter is fairly common in the software industry.

In psychological literature, time pressure refers to situations where time is a limited resource ~\cite{maule2000effects}. There are several well-validated theories related to time pressure and its effects on stress, decision making, and motivation, such as the Yerkes-Dodson law~\cite{yerkes1908relation}, the job demands-resources model~\cite{bakker2007job}, and the speed-accuracy trade-off~\cite{heitz2014speed}. 
These theories, which we discuss in more detail in Section ~\ref{sec:general}, are relevant to time pressure in software engineering, despite having been developed in other fields.

Project overruns and overtime work in the software engineering industry are reported as common by both academic~\cite{molokken2003review} and practitioner~\cite{computerworld,kotaku} sources. Yet, we argue that a systematic review of the current understanding of time pressure and its effects on productivity in software engineering is needed, in order to provide a more solid basis for future research. For example, by compiling together previously used metrics of time pressure, our work enables future studies to conduct well-informed measurements of time pressure in software engineering.

This review heavily extends our previous work~\cite{kuutila2017reviewing}, in which clustering on Scopus data was performed to identify in literature more specific topics on time pressure in software engineering and related disciplines. We partly use that work to establish a ``seed" set of papers, which we complement with novel, relevant articles, with Google Scholar searches. The papers are expanded using Wohlin's snowballing guidelines~\cite{wohlin2014guidelines}. Our goal is to provide an overview of the existing literature related to time pressure in software engineering and map it to different process phases, as well as to synthesize the gathered information in a way that provides new information. The latter is accomplished by seeking answers to the following research questions:
\begin{description}
\item[RQ1-Definitions] {\RQOne}
\item[RQ2-Metrics] {\RQTwo}
\item[RQ3-Process Phases] {\RQThree}
\item[RQ4-Causes] {\RQFour}
\item[RQ5-Effects and Outcomes] {\RQFive}
\end{description}

The goals of a systematic mapping study are to provide an overview of a research topic, identify relevant quantities and type of research~\cite{petersen2008systematic}, while the goal of a systematic literature review aims to summarize and examine to what extent does empirical evidence support or contradict the considered hypotheses~\cite{kitchenham2007guidelines}. Hence, the first four research questions are more related to systematic mapping studies, while \emph{RQ5-Effects and Outcomes} is applicable to systematic literature reviews. Although our paper combines elements of both, we title our work a systematic review, for convenience. The remainder of this article is structured as follows. In Section~\ref{sec:general}, we outline background information on theories and concepts related to time pressure. In Section~\ref{sec:methodology}, we introduce the methodology used in this study. In Section~\ref{sec:definition}, we present the definitions of time pressure, discuss the different metrics that have been used, and elaborate on previous theoretical work. In Subsections~\ref{sec:definitions} and~\ref{sec:metrics}, research questions \emph{RQ1-Definitions} and \emph{RQ2-Metrics} are answered, respectively. In Section~\ref{sec:phase}, we map a number of different studies to processes and approaches, to answer \emph{RQ3-Process Phases}. Section~\ref{sec:empiricalresults} contains an overview of the existing literature and summarizes the empirical descriptives provided by several studies. In Sections~\ref{sec:causes} and~\ref{sec:summary}, research questions \emph{RQ4-Causes} and \emph{RQ5-Effects and Outcomes} are answered, respectively. Last, in Section~\ref{sec:conclusions}, we conclude the paper by outlining our contributions and providing a series of takeaways for practitioners and researchers.
\section{Related Work}\label{sec:general}
In occupational and social psychology, time is considered to be a resource. The scarcity of resources, such as time and money, and the effects of scarcity on mindset and human perception have been a subject of popular science ~\cite{mullainathan2013scarcity}. Mullainathan and Sharif~\cite{mullainathan2013scarcity} make the case that scarcity of time introduces both {\it tunneling} and {\it focus} mindsets. Tunneling, in this context, means valuing short-term over long-term goals that are related to the scarcity of a resource. An example could be when time-scarce software engineers prefer a solution that is quick to implement, regardless of its impact on the longevity of the software. However, a scarcity mindset can also make individuals focus when spending a limited resource. In our example, scarcity of time would make software engineers avoid gold plating, i.e. working on a task beyond what is reasonably expected. In many studies, time pressure is defined as the perception that time is {\it scarce} in relation to the demands of the task~\cite{basten2017role, kelly1985effects, cooper2001organizational}. 

\begin{figure}[htp]
\caption{The Yerkes-Dodson law}
\includegraphics[width=1.0\textwidth]{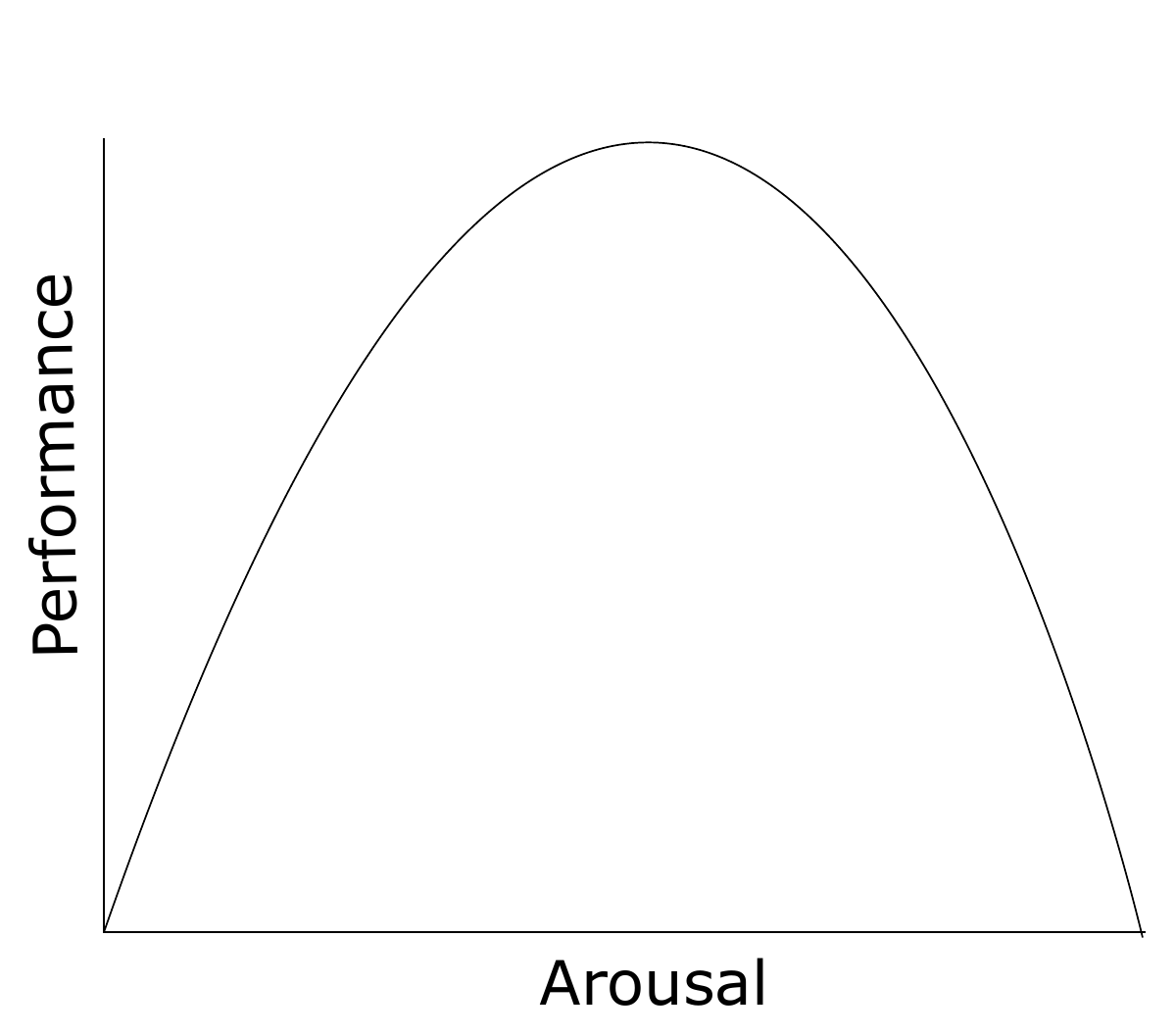}
\label{fig:yerkes}
\end{figure}

Another view on time pressure comes from the Yerkes-Dodson law, which dates back to the start of the 20th century~\cite{yerkes1908relation}, and states that arousal, caused by (time) pressure, and performance have an inverted U-shaped relationship. This is depicted in Figure \ref{fig:yerkes}. In other words, arousal increases performance, but only up to a certain point, from where performance starts decreasing. In this view, time pressure is seen to increase the activation level and urgency (arousal).

The speed-accuracy trade-off~\cite{heitz2014speed} offers another lens for viewing time pressure: it is observed that the decision speed negatively correlates with the quality of the decision. In fact, the phenomenon is ubiquitously observed, with species such as ants or bumblebees~\cite{heitz2014speed}. The speed-accuracy trade-off is commonly used as a benchmark for decision processes across task-domains, as simply examining either reaction speed or quality alone is not a sufficient benchmark. The speed-accuracy trade-off has also been, in part, the subject of human computer-interaction studies~\cite{mackenzie2008fitts}.

In occupational psychology, the well-known job demands-resources model~\cite{bakker2007job} is used to explain employee well-being. Generally, the model assumes every job to have both demands and resources, and well-being to be the result of their balance. Resources in the model refer to skills, autonomy, feedback, and others, while demands can include role ambiguity, performance, and emotional demands. Hence, time pressure can be seen as a demand, with too much of it leading to worse well-being outcomes, exhaustion, and even burnout~\cite{demerouti2001job}.

Evidence has also shown teams may work exceptionally well under extreme time pressure, such as the Apollo 13 ground crew ~\cite{chong2011double}. Thus, in a work by \citet{chong2011double}, a challenge-hindrance framework~\cite{lepine2005meta} for time pressure was deemed appropriate. In this view, time pressure is defined as having either positive (challenge) or negative (hindrance) effects on goal achievement. \citet{lepine2005meta} pointed out that challenges could be viewed as good stress, while hindrances as bad stress. Thus, not only the amount, but also the type of time pressure matters. Examples from \citet{chong2011double} of hindrance (bad) time pressure are  ``amount of constant switching between tasks'' or ``impossibility to fulfill the project schedule,'' while challenge (good) time pressure item examples are  ``importance of completing this project on time'' or ``urgent need for successful completion of the work the team is doing.'' As recognized by \citet{chong2011double}, the boundaries between challenge and hindrance are not always clear, and \citet{chong2011double}'s survey had several items that did not clearly fall in either category. In software engineering, the challenge-hindrance time pressure definition has been used by \citet{lohan2014investigation}.

As demonstrated by the clustering performed in our previous work ~\cite{kuutila2017reviewing}, time pressure has been studied across different occupations and fields, e.g., accounting, medicine, construction work, and vessel operating pilots. The mediating role of knowledge under time pressure haven been previously suggested: for example, accounting professionals (considered a high-knowledge group) improved their performance under time pressure, whereas the performance of students (low-knowledge group) decreased~\cite{spilker1995effects}. Similarly, industry-specific auditors have been reported to experience less time pressure than non-specialized auditors~\cite{huang2015experimental}. In nursing, time pressure has been linked to worse patient safety, but only together with high levels of burnout~\cite{teng2010interactive}. Another context where time pressure has been studied is related to risky choice behaviour, the majority of the studies indicating higher risk taking under time pressure~\cite{madan2015rapid}.
\section{Methodology}\label{sec:methodology}

\begin{figure}[htp]
\caption{Flow chart of our research methodology.}
\includegraphics[width=1.0\textwidth]{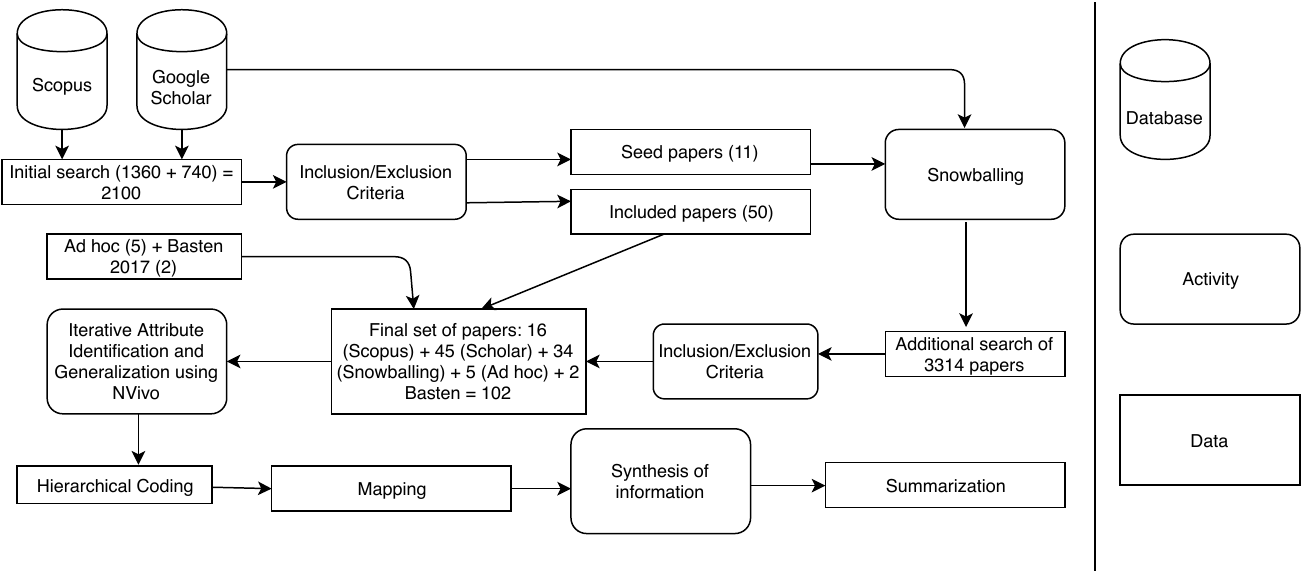}
\label{fig:flowchart}
\end{figure}

In this section, we present the methodology used in this study. Subsection~\ref{sec:criteria} introduces the inclusion and exclusion criteria we used to grade the found literature. In Subsection~\ref{sec:database}, we present the database search strings and approaches. In Subsection~\ref{sec:snowballing}, we explain the snowballing procedure in detail. Last, in Subsection~\ref{sec:dataextraction}, we explain the data extraction in detail. The entire research methodology is shown in Figure~\ref{fig:flowchart}.

\subsection{Selection criteria and selection process}\label{sec:criteria}

As previously stated, our goal is to provide an overview of time pressure in the software engineering context, that overviews the definitions, metrics, process phases, causes and effects found in the literature. This goal leads to our first inclusion rule (I1).

The second inclusion rule (I2) is about studies that include time pressure variables and provide empirical evidence, but whose main focus might lie elsewhere. For example, integration failures~\cite{cataldo2011factors}. The second inclusion rule gives this paper a broad scope and highlights less well-known papers related to time pressure in software engineering that are not as easily found. In practice, we excluded papers with brief mentions of deadlines or time pressure, which were not derived based on a systematic effort to gather information, as anecdotal level evidence. For example, sources where time pressure is used to explain unexpected results ex post facto as a possible contributing factor. We define empirical evidence as either quantitative studies conducted with data including variables related time pressure, or qualitative studies which study and have practical takeaways in time pressure and time usage related issues. Hence our inclusion criteria takes into account quality of evidence.

We used the following inclusion criteria:
\begin{description}
\item[I1]
The main focus is time pressure in software engineering.
\item[I2]
The paper presents empirical evidence of time pressure in software engineering.
\end{description}

We formed the first and second exclusion rules (E1 and E2) at the beginning of the search to quickly exclude sources that we could not reliably interpret, or which were not published in scientific venues. The exclusion rules E3 was formed iteratively throughout the literature search process when we encountered studies that were clearly related to time pressure and computers, but that did not relate directly to software development, such as studies on the end users of information systems. We used the following exclusion criteria:
\begin{description}
\item[E1]
The paper was written in language other than English.
\item[E2]
Not a scientific source.
\item[E3]
The task studied is not a software development task.
\end{description}

Only one of the inclusion criteria had to be present in a paper for it to be included. In practice, this meant that papers without empirical evidence had less chance to be included, as their focus had to be time pressure to be included with inclusion rule I1. One such example is the agency model by Austin~\cite{austin2001effects}, which only satisfies the inclusion rule I1. However, multiple papers we included fulfill both rules of the inclusion criterion.

The use of exclusion criteria differed based on the rule. Rules E1 and E2 excluded a paper as soon as we observed the paper was written in a language other than English or that it was not published in a scientific venue. However, when we discovered papers with elements of exclusion criterion E3, we examined them so that we could come to a verdict. In other words, if the papers made contributions related to I2, but all these contributions were covered by exclusion criteria E3, the paper was excluded. However, if the paper made some contributions related to inclusion criteria I2, for which exclusion criterion E3 was not related, we included the paper.

The selection process for each paper started with the first author reading the title and the abstract of the paper. In cases where no definitive decision could be made based on these elements, the first author read the conclusions and other relevant sections of the paper until we could make a decision based on the selection criteria. The first author was advised to default to caution when unclear cases were found, as we did not want to miss any potential papers. This made the selection process more laborious, but reading just the abstract would not have uncovered papers making significant and relevant contributions such as \citet{cataldo2011factors}. 

Unclear or borderline cases were marked down and discussed by the first two authors until a decision for inclusion or exclusion could be made. We used two person review only for borderline papers. We settled upon this approach as we had very large set of papers to cover and as most of the exclusion decisions were simple, as can be seen in Table \ref{tab:exclusion}. This meant that we could afford careful pondering on the borderline cases. For transparency, also all borderline cases are marked in our \emph{replication package}\footnote{\url{https://figshare.com/s/0662c66e0705ebf8dca7}}. A total of 129 papers were marked unclear and discussed by the two first authors. Of these 129 papers, we included 23 in the literature review. Occasionally, the unclear cases led to clarification of the interpretation of the inclusion and exclusion rules. For example, we did not anticipate that there would be papers from non-software engineering forums that gave in-depth narrative descriptions of software companies with extensive time pressure and its consequences, such as~\citet{borg2014ciar}, who observed an ICT company and the problems arising from time pressure. The first author went through all the sources found with database searches twice. This was to ensure our interpretation of the rules were consistently enforced, additionally we had not explicitly recorded the reasons for inclusion and exclusion during the first round.

\begin{table}[]
\caption{Search strings, total results and included papers from the database search}
\label{tab:search}
\begin{adjustbox}{width=\textwidth}
\begin{tabular}{l|l|l|l}
\textbf{Search Engine}  & \textbf{Search String}                     & \textbf{Results} & \textbf{Included} \\
\hline
Scopus         & Title or keyword: ``time pressure'', ``schedule pressure''&&\\
& ``time budget pressure'', ``deadline pressure''&& \\ 
& ``pressure of time'', ``pressure of schedule''&& \\ 
& ``pressure of time budget'', ``pressure of deadline''&& \\ 
& ``speed accuracy tradeoff''&&\\
&And not title, abstract, or keyword:&& \\
& ``long rise-time'', ``intracranial'', ``drill'', ``space-time'' & 1270    & 12        \\
Scopus & ``schedule compression'' AND ``software engineering''           & 91     & 4       \\
Google Scholar & ``time pressure'' AND ``software engineering''           & 100     & 13       \\
Google Scholar & ``schedule pressure'' AND ``software engineering''       & 100     & 16       \\
Google Scholar & ``time budget pressure'' AND ``software engineering''   & 100     & 3        \\
Google Scholar & ``deadline pressure'' AND ``software engineering''       & 100     & 11       \\
Google Scholar & ``pressure of time'' AND ``software engineering''       & 100     & 0        \\
Google Scholar & ``pressure of schedule'' AND ``software engineering''    & 19      & 0        \\
Google Scholar & ``pressure of time budget'' AND ``software engineering''  & 4       & 0        \\
Google Scholar & ``pressure of deadline'' AND ``software engineering''   & 17      & 1        \\
Google Scholar & ``speed-accuracy tradeoff'' AND ``software engineering'' & 100     & 0       \\
Google Scholar & ``schedule compression'' AND ``software engineering'' & 100     & 1       \\
\end{tabular}
\end{adjustbox}
\end{table}

\subsection{Database search methodology}\label{sec:database}

Part of the initial database search was based on our previous work~\cite{kuutila2017reviewing}, in which we used the Scopus database search with Latent Dirichlet Allocation (LDA) clustering to get an overview of where time pressure has been studied. As advocated by Kitchenham and Charters~\cite{kitchenham2007guidelines}, we used multiple trial searches to establish keywords and synonyms for time pressure, all of which can be found in Table~\ref{tab:search}. We used Scopus and Google Scholar for the primary literature search from databases. All used search strings can be found in Table~\ref{tab:search}, together with the number of total results and the number of primary sources we included for our review. Because Google Scholar also searches in the article full-text, while Scopus searches only for title, abstract, and keywords, we used the tool \emph{Publish or perish 5}\footnote{\url{https://harzing.com/resources/publish-or-perish}} to complement the search from Scopus with Google Scholar searches. For the complementary Google Scholar search, we took only the first 100 results for each search. Additionally, we performed some ad hoc searches, and they resulted in six additional papers. Altogether, we included 61 sources from the 2,100 found with repository searches. In the flowchart described in Figure~\ref{fig:flowchart}, the initial search can be found in the activity "Initial search," and the additional papers found with ad hoc searches are added to the final set of papers as "+7."

\subsection{Snowballing}\label{sec:snowballing}
Wohlin's snowballing guidelines~\cite{wohlin2014guidelines} were used to construct a set of seed papers and snowball through them. The set of papers used for snowballing contained papers we found with repository searches (Section~\ref{sec:database}). We included 11 papers, with broad range of authors and publication fields and venues. One round of snowballing was performed forward and backward for these papers. Additionally, all publications published by the authors included in the seed paper set were examined with the selection criteria. We used only Google Scholar to identify papers needed for the snowballing. Indeed, we had determined that effort was better spent on reviewing more papers on the snowballing phase rather than double checking two search engines (Scopus and Google Scholar) for a smaller number of additional papers.

In the snowballing phase, 3,314 papers were evaluated based on the selection criteria, and it resulted in the inclusion of 34 additional papers. The snowballing phase had an inclusion rate of 1.03\%.

Before we started hierarchical coding, two of the authors went through all the included papers together and discussed their relevance for the topic. Based on these discussions we excluded six papers, as in practice these papers included only anecdotal evidence. These six exclusions are not reported in the previous numbers.

While we were conducting this literature review, work by \citet{basten2017role} has been published. Our work at that stage already included 9 of the 13 papers Basten introduced in his work. Of the four not included, based on our inclusion and exclusion criteria, we decided to include two, as well as the paper by Basten. This explains the +2 in Figure~\ref{fig:flowchart} in the final set of papers. Overall, our work is broader than Basten's as we have 102 papers while Basten had 13 of which we include all but two.

\begin{table}[]
\caption{Verdicts}
\label{tab:exclusion}
\begin{adjustbox}{width=\textwidth}
\begin{tabular}{l|l|l|l|l}
Verdict & Scopus & Scholar & Snowballing & \% \\
\hline
The main focus is time pressure in & & & & \\
software engineering (I1) & 10 & 5 & 11 & 0.5\%\\
\hline
The paper presents empirical evidence of & & & & \\
time pressure in software engineering (I2) & 6 & 40 & 23 & 1.3\%\\
\hline
Already Included, duplicate & 4 & 35 & 81 & 2.2\%\\
\hline
No empirical evidence on time pressure, & & & & \\
not focused on time pressure (I1 \& I2)  &  73 & 523 & 1366 & 36.2\%\\
\hline
Not a software development paper (E3)  &   1246     &  109  &  1491   & 52.6\%     \\
\hline
Not a scientific source (E2)& 15 & 10 & 19 & 0.8\% \\
\hline
Not available & 7 & 17 & 194 & 4\%\\
\hline
In language other than English (E1) & & 1 & 128 & 2.4\% \\
\hline
\end{tabular}
\end{adjustbox}
\end{table}

We have also provided verdicts for all papers in Table~\ref{tab:exclusion}. This table only includes verdicts for papers found with the database search strategy and snowballing. This means that papers found Ad hoc (see figure ~\ref{fig:flowchart}), such as \citet{basten2017role}, are not included in the table.

As can be seen in Table~\ref{tab:exclusion}, the most common reason to exclude a paper was when its context was deemed outside of software engineering. This, for example, included papers focusing on users of information systems rather than developers. The second most common reason to exclude a paper was not meeting any of inclusion criteria. Not available verdict includes sources with incomplete bibliographic information, non-digitalized material, as well as some papers behind paywalls we did not have access to. Of the included papers, 29 were included with the inclusion criteria I1 and 73 were included with the criteria I2.

\subsection{Data extraction}\label{sec:dataextraction}
The papers we included were qualitatively coded and analyzed using QSR International's NVivo\footnote{https://www.qsrinternational.com/nvivo/home}. We followed~\citet{zhang2016qualitative} qualitative guidelines. The coding scheme was both used to familiarize ourselves with the material, as well as to form the basis for Sections~\ref{sec:definition}, \ref{sec:phase} and \ref{sec:empiricalresults}.

Coding started with few ready codes, such as process phases and causes of time pressure, but it was iteratively improved. In the end, highest level of the coding included classes empirical results, definitions, research methodology, research questions and hypotheses, and lastly process phases. The coding for process phases formed the basis for the mapping in Section \ref{sec:phase}, while coding for empirical results formed the basis for Section \ref{sec:empiricalresults}. In the first round, the first author coded all the sources. The resulting coding scheme was improved and checked by the second author. Once every paper was coded and the coding scheme was ready, the first author went through all the papers a second time to apply the scheme consistently on every source. Finally, all the authors checked the results and analyzed a part of the coding that was given to them. In practice, this meant that at least two persons checked all parts of the coding scheme and coded text. We have included a picture depicting the coding scheme in our \emph{replication package}\footnote{\url{https://figshare.com/s/0662c66e0705ebf8dca7}}.
\section{Definition, metrics, and previous theoretical work in software engineering}\label{sec:definition}
In this section, we provide and examine the definitions given for time pressure in the previous literature, summarize the metrics used to measure it, and introduce prior theoretical works focusing on time pressure in software engineering context. Research question \emph{RQ1-Definitions} about which definitions of time pressure are used is answered in Subsection~\ref{sec:definitions}. Additionally, research question \emph{RQ2-Metrics} about which metrics are used to measure time pressure is answered in Subsection~\ref{sec:metrics}.

\subsection{Definition}\label{sec:definitions}

Multiple synonyms for time pressure exist in the scientific literature about software engineering, such as schedule pressure~\cite{ebert2009embedded}, deadline pressure~\cite{costello1984software}, and time budget pressure~\cite{nan2009impact}. Schedule pressure and deadline pressure emphasize a deadline or deadlines when a task or project should be done while time budget pressure highlights the amount of time that can be used for a task or a project.

Pressure due to compression of the schedule was considered in the early work of software engineering. Barry Boehm~\cite{boehm1981software} defined schedule compression as the percentage of schedule cut in a project's planned duration compared to the nominal schedule of the project. This a definition was used later in research into project simulation models~\cite{jeffery1987time, abdel1990investigating}.

\citet{powell1999strategies} defined schedule pressure as the relationship between required and applied productivity. Additionally, they discovered that ``it is possible to increase pressure on the development team (by requiring additional productivity) but only up to a certain point, after which productivity rapidly declines,'' pointing at the Yerkes-Dodson law~\cite{yerkes1908relation}.

To summarize, the definitions of time pressure focus on an individual's perception of the time scarcity~\cite{basten2017role, kelly1985effects, cooper2001organizational}, or on the project level of schedule compression~\cite{boehm1981software,powell1999strategies}. The two main concepts of how time pressure is understood in software engineering literature are the U-shaped relationship between arousal and performance~\cite{yerkes1908relation} and the division of time pressure into challenge and hindrance time pressure~\cite{chong2011double}.

\subsection{Metrics and operationalization}\label{sec:metrics}
Gathering metrics for time pressure is essential. As different definitions of time pressure exist, metrics for operationalization of those definitions reveal more details about time pressure. The metrics in empirical settings can vary from study to study, and they also depend on what is available in each context. Summarizing the metrics helps future researchers as they can consider the existing metrics when designing their studies. Table~\ref{tab:metrics} shows all the metrics we found to measure time pressure.

Questionnaire and survey-based metrics that use an ordinal scale to measure time pressure are popular~\cite{park2008overcoming}. Early on, \citet{banker1987factors} simply asked project leaders if the deadline pressure was higher than average. Similarly, \citet{mukhopadhyay1992software} asked to rate software projects on a scale of one to four regarding deadline pressure, where four signaled very high pressure. \citet{durham2000effects} developed a scale specifically designed to measure time pressure, which \citet{maruping2015folding} used to study time pressure in software projects. General questionnaires, such as the NASA Task Load Index (NASA-TLX), have a question about the temporal demand of a task and have been used to measure time pressure in software engineering~\cite{mantyla2014time}. 

The literature of software cost estimation uses metrics-based estimated effort. \citet{ruiz2001simplified} defined schedule pressure as estimated effort, minus the remaining effort divided by the estimated effort, meaning a positive value indicates delayed project, while a negative value indicates a project that is advancing according to the initial estimates. \citet{nan2003impact} defined time pressure as the estimated time for the project minus the customer negotiated time for the project divided by the estimated time for the project. This work is similar to \citet{ruiz2001simplified}, but the difference is that they used customer-negotiated time instead of estimated effort. Cycle time was defined as the project duration starting from the first day of design work and continuing until the customer accepts the delivered product. The authors also used this metric in a more well-known later work from 2009~\cite{nan2009impact}. In COCOMO II~\cite{boehm2000cost} actual schedule compression is defined as the ratio between actual schedule and the estimated nominal schedule.

\citet{cataldo2010sources} used a metric, called \emph{the task temporal metric}, to estimate time pressure experienced in a project. It is calculated as the standard deviation of modification requests completed each month. The author contented that the high value of the task temporal metric is associated with an uneven workload during the project which suggests time pressure in the months with a high number of tasks. 

More recently, there have been efforts to detect time pressure with sentiment analysis and various sensors. \citet{kolakowska2013emotion} introduced a multi-modal emotion recognition application, which combined physiological, video, and depth sensors, to train a classifier to be used with several software engineering methods. The motivation for the work in part came from future work of detecting stress induced by time pressure and an investigation of productivity and emotions. Similarly, using sentiment analysis, \citet{mantyla2016mining} found that higher arousal (e.g., activation level) was associated with more severe issue reports. In a later work, \citet{mantyla2017bootstrapping} developed a lexicon for sentiment analysis for more efficient detection of arousal levels in the software engineering context.

\begin{table}[]
\centering
\caption{Metrics identified from the collected literature}
\label{tab:metrics}
\begin{tabular}{l|l|l}
\textbf{Metric to Measure Time Pressure}          & \textbf{Example Papers}   & \textbf{Data Source}                                                                                                \\
\hline
$\frac{\mbox{Estimated time } - \mbox{ customer negotiated time}}{\mbox{estimated time}}$    &&                        \\ 
& ~\cite{nan2003impact}  & Company Project Database \\
 & ~\cite{nan2009impact}& \\
\hline
$\frac{\mbox{Actual Schedule}}{\mbox{Nominal Schedule}}$ & \cite{boehm2000cost} & Company Project Database\\
& & \\
\hline
$\frac{\mbox{Estimated effort } - \mbox{ Remaining effort}}{\mbox{Remaining effort}}$ & & \\
& \cite{ruiz2001simplified} & Model simplification \\
\hline
Standard deviation of tasks &                                                              \\ completed in a project each month
& \cite{cataldo2010sources}   & Company Project Databases \\
\hline
Questionnaires and surveys                                    & \cite{park2008overcoming}    & Questionnaires and surveys   \\
& \cite{maruping2015folding}&  \\
\hline
Physiological measurements                                    & \cite{kolakowska2013emotion} & Skin Conductance \\  					 &~\cite{tuomivaara2017short} & Electromyography\\
&& Electrocardiography, etc.\\
\hline
Sentiment analysis                                            & \cite{mantyla2016mining, mantyla2017bootstrapping}  & Natural Language Text
\end{tabular}
\end{table}

We present experimental designs used to create time pressure; see Table~\ref{tab:experiments}. They are not about measuring time pressure but represent essential information on the operationalization of time pressure. An early paper by \citet{hwang1994decision} noted that time pressure can be operationalized as the time available for task performance, for example, as different time limits in experiment settings. Different time limits~\cite{juristo2011role} and task difficulties~\cite{ramanujan2000experimental} have been widely used in experimental settings as operationalizations of time pressure. Rewarding faster completion is another alternative to create time pressure in experiments~\cite{mantyla2014time}.

\begin{table}[]
\centering
\caption{Creating time pressure in Experiments}
\label{tab:experiments}
\begin{tabular}{l|l|l}
\textbf{Creating Time Pressure}          & \textbf{Example Papers}   & \textbf{Data Source}                                                                                                \\
Time limits in                          & \cite{juristo2011role} & performance in the experiment                                                    \\
experimental settings& ~\cite{mantyla2013more}& \\
\hline
Task difficulties in                    & \cite{ramanujan2000experimental} & performance in the experiment \\
experimental settings& & \\
\hline
Reward for faster completion                                    & \cite{mantyla2014time}  & performance in the experiment                \\
in experimental setting&& \\
\end{tabular}
\end{table}

\subsection{Theoretical papers and reviews}
During our search for academic literature, we did not find any previous sources following systematic literature review guidelines to assess previous work related to time pressure in software engineering. However, we found theoretical work and non-systematic literature reviews that focused mainly on time pressure in software engineering.

Early focus on time pressure in software engineering is related to cost models and cost estimation. \citeauthor{costello1984software}'s paper from \citeyear{costello1984software} is a prime example of a purely theoretical paper. The paper presents a simplistic scheduling model and discusses schedule pressure at length based on experiences. The main contribution of the paper is a list of resource allocation strategies aimed at decreasing the effects of schedule pressure.

Widely cited paper by \citet{austin2001effects} presents an agency framework focused on the effects of time pressure on software quality. Based on the modeled framework, the author recommends setting aggressive deadlines, where it is okay to miss deadlines. The author also concluded that adding slack time does not necessarily minimize costs and that deadlines should be set separate from planning estimates.

\citet{malgonde2014applying} presents a research proposal about how emergent outcome controls are adapted when time pressure increases. The authors planned to investigate with interviews and critical incident method. 

\citet{harris2009agile} considers time-related concepts and issues in Agile software development and introduces research propositions for the future. These concepts and issues are highlighted with data from qualitative interviews. The authors compare the role of deadlines in traditional software projects that use the waterfall life cycle and software projects that use Agile methods. The authors argue that employee motivation and stress should be compared with multiple shorter deadlines versus one longer deadline in the Agile software development context, and additionally comparing time-to-completion with higher and lower uncertainty projects developed with Agile and plan-driven approaches.

\citet{basten2017role} present a literature review and a research agenda basis for future studies. Part of the research agenda is a call for methodological pluralism, as the author argued that previous works did not use qualitative research methods. \citet{basten2017role}) also argue for research agenda conceptualization (e.g., better definitions of time pressure), research on contemporary development approaches, such as Agile and Scrum, better definitions of the role of the context of time pressure to better understand the diverse results, and empirical validation in the form of replication studies.

Last, our previous work~\cite{kuutila2017reviewing} presented a computer-aided literature review and introduced testable hypotheses related to time pressure from fields other than software engineering. The paper presented a list of testable hypotheses in software engineering related to time pressure derived from fields other than software engineering. For example, under time pressure, fewer test can run and hence less feedback is provided, or more bugs are introduced to the code in more complex classes near deadlines. Additionally, that paper formed the foundation for this paper.

\section{Mapping}\label{sec:mapping}

In this section, we provide bibliographical information the selected studies,  map them to process phases and methodologies, as well as to used research methods. In Subsection \ref{sec:bio}, we provide bibliographical information for the selected studies and their mapping to the adopted research methods. Next, Subsection \ref{sec:phase} discusses the mapping to process phases and methodologies.

\subsection{Publication years, venues and used research methods}\label{sec:bio}

\begin{figure}[htp]
\caption{Number of publications per year.}
\includegraphics[width=1.0\textwidth]{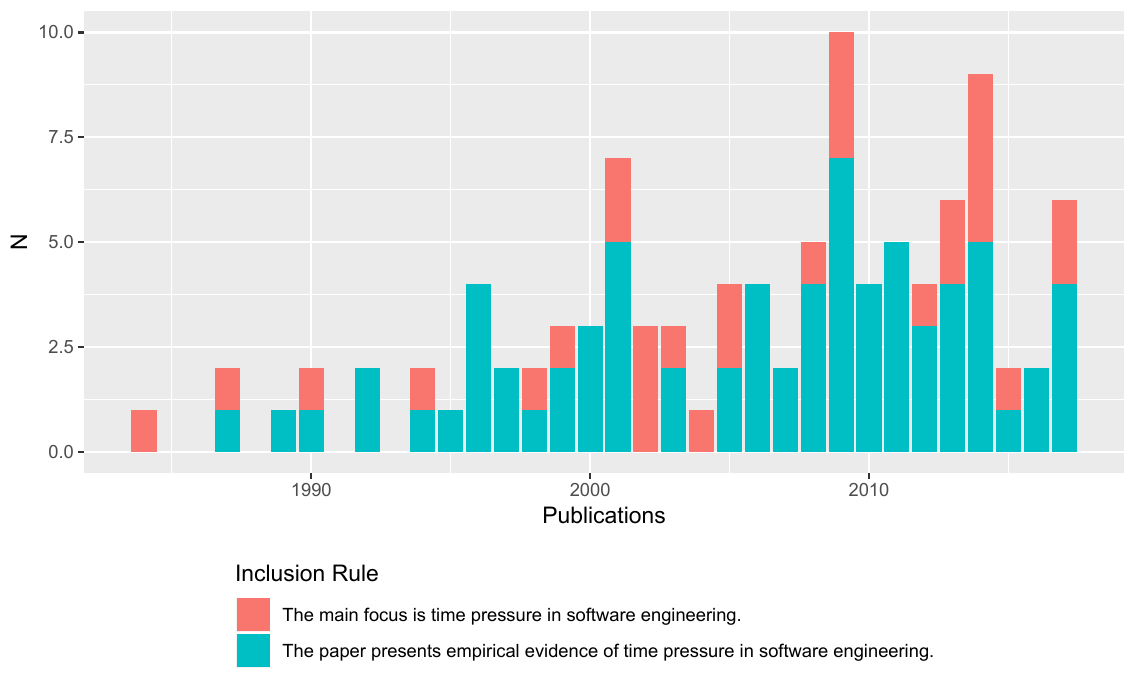}
\label{fig:years}
\end{figure}

Figure~\ref{fig:years} shows the publication years of all the papers included in the present study. It can be observed that the number of publications is larger after the year 2000, than before it. All the earliest publications focus on cost estimation and process simulation models, e.g., \citet{costello1984software} and \citet{jeffery1987time}. More recent papers include Agile methodologies (e.g., \citet{laanti2013agile}), software process improvement (e.g., \citet{agrawal2007software}), and time pressure detection (e.g., \citet{kolakowska2013emotion}). 

Similarly, Table~\ref{tab:publications} shows the most common publication venues, which published two or more papers of the articles we included in the study. We note that 35 were published in conferences or workshops, 57 in academic journals, 7 in academic magazines (e.g., IEEE Software, Computer, Communications of the ACM, etc.) and 2 in PhD theses.

\begin{table}[t]
\centering
\caption{Most common publications venues}
\label{tab:publications}
\begin{tabular}{l|l}
Publication & N \\
\hline
Journal of Systems and Software & 10 \\
Information and Software Technology & 7 \\
International Conference on Software Engineering (ICSE) & 5 \\
IEEE Transactions on Software Engineering & 5 \\
Americas Conference on Information Systems (AMCIS) & 4 \\ 
Academy of Management Journal & 3 \\ 
Computer & 3 \\
International Symposium on Empirical Software Engineering and Measurement (ESEM) & 3 \\
Journal of Management Information Systems & 3 \\
Communications of the ACM & 2 \\
CrossTalk - The Journal of Defense Software & 2 \\
Hawaii International Conference on System Sciences (HICSS) & 2 \\
IEEE Software & 2 \\
Information Systems Research & 2 \\
International Conference on Information Systems (ICIS) & 2 \\
International Conference on Mining Software Repositories (MSR) & 2 \\
International Journal of Human-Computer Interaction & 2 \\
Others & 43 \\
\end{tabular}
\end{table}

We have mapped the included studies also according to the adopted research methodologies. We define as "Theoretical Papers \& Reviews" (see Table \ref{tab:methodology}) articles that do not provide their own original empirical evidence (i.e., secondary and tertiary studies). Next, position pieces include positional papers arguing for a position or a research proposal, but lacking experimentation and original research. We have defined studies that create and evaluate an artifact (e.g., a simulation model) as design science research \citet{peffers2007design}. Otherwise, for empirical research we followed the categorization by Easterbrook et al. \citet{easterbrook2008selecting}, with the addition of a ``company data"  category for studies that quantitatively analyze software project data from multiple projects. We present the result in Table~\ref{tab:methodology}. The most commonly used research method is Design Science. The code to reproduce Tables ~\ref{tab:publications} and \ref{tab:methodology} and Figure \ref{fig:years} is given in our \emph{replication package.}\footnote{\url{https://figshare.com/s/0662c66e0705ebf8dca7}}

\begin{table}[t]
\centering
\caption{Number of implemented research methodologies.}
\label{tab:methodology}
\begin{tabular}{l|l}
Methodology & N \\
\hline
Design Science Research & 19 \\
Mixed Methods & 14 \\
Survey & 13 \\
Experiment & 12 \\
Case Study & 12 \\
Company Data & 12 \\
Theoretical Papers \& Reviews & 9 \\
Position Piece & 7 \\
Ethnographic Study & 3 \\
Action Research & 1 \\

\end{tabular}
\end{table}

\subsection{Process phase or approach}\label{sec:phase}

\begin{table}[!htp]
\caption{Found papers by investigated software development process phases and methodologies}
\label{tab:phase}
\begin{tabular}{p{7em}|p{9em}|p{13em}}
\hline
\textbf{Category} &\textbf{Sub-Category} & \textbf{Papers} \\
\hline
Process phase &Requirements engineering
&
~\cite{bjarnason2012you, ferreira2009understanding}
\\
\hline
Process phase& Design and Acquisition &
~\cite{fehrenbacher2014behavioural, hwang1994decision, nugroho2007survey, rosenberg2017rapid, tang2006survey}
\\
\hline
Process phase&Programming and implementation & 
~\cite{marques2010coordination, plonka2012disengagement, sanjram2013task, sojer2014understanding, topi2005effects}
\\
\hline
Process phase&Quality Assurance &
~\cite{baskerville2001internet, cataldo2010sources, cataldo2011factors, ciolkowski2003software, deak2016challenges, ebert2009embedded, hale2011evaluation, juristo2011role, karhu2009investigating, lavallee2015good, leszak2000case, mantyla2016mining, mantyla2013more, mantyla2013rapid, mantyla2014time, shah2014global} \\
\hline
Whole Process or approach &Evolution and maintenance:  & 
~\cite{banker1987factors,hale2011evaluation, ramanujan2000experimental}
\\
\hline
Whole Process or approach&Agile and Scrum: &  
~\cite{bjarnason2012you, harris2009agile, laanti2013agile, lohan2014investigation, mann2005case, riordan2013time, rosenberg2017rapid, tuomivaara2017short, van2009dynamics} \\
\hline
Whole process or approach&Process improvement & ~\cite{agrawal2007software, baddoo2003motivators, baddoo2000implementing, niazi2009software, niazi2005maturity, paulish1994case, smite2009cmmi}
\\
\hline
Whole process or approach &Cost estimation, cost models, simulation and project escalation: & ~\cite{abdel1990investigating, abdel1989lessons, abdel1990elusive, basten2011current, bjarnason2012you, cao2010modeling, costello1984software, houston2001stochastic, hu1998software, hughes1996expert, hussain2008effect, jeffery1987time, kitchenham1992empirical, lee2012effect, lin1997software, miranda2008protecting, mukhopadhyay1992software, nan2003impact, nan2009impact, pfahl2001integrated, powell1999strategies,  rahmandad2009dynamics, reichelt1999dynamics, rosenberg2017rapid, ruiz2001simplified, smith2001empirical, yang2008survey, yang2005effect, zhang2006qualitative}\\
\hline
Whole Process or Approach&Project Success and Failure& ~\cite{ebert2009embedded, jones2006social, nasir2011critical, procaccino2006defining, verner2008factors}\\
\hline
Other&Detection of time pressure &
~\cite{kolakowska2013emotion, mantyla2016mining, mantyla2017bootstrapping}
\\
\hline
Other&Group interaction & ~\cite{langer2014project, lohan2014investigation, marques2010coordination, maruping2015folding, park2008overcoming}
\\
\hline
Other&Fields other than Software Engineering and Information Systems& ~\cite{borg2014ciar, clegg1996software, fujigaki1996time, fujigaki1997longitudinal, jemielniak2009time, perlow1998boundary, perlow1999time, perlow2001time, perlow2002speed, sawyer2002temporal, Staudenmayer2002, tapia2004power,  wilson2014testing} \\
\hline
Other&Literature reviews and theoretical papers: &  
~\cite{austin2001effects, basten2017role, costello1984software, harris2009agile, kuutila2017reviewing, malgonde2014applying}
\\
\hline
Other&New product development & ~\cite{blackburn1996time, van2013anatomy, van2010get}
\\
\hline
Other& Individual psychological factors & ~\cite{ fujigaki1996time, fujigaki1997longitudinal, graziotin2017unhappiness, laanti2013agile, paul2012time, tuomivaara2017short, turley1995competencies}
\\
\hline
\end{tabular}
\end{table}

In this section, we summarize and map the papers we found into different process phases and methodologies, to give the context of their results. This can be useful for a scientist who is interested in a particular process area. We map the papers to the different process phases of the waterfall model~\cite{sommerville1996software}. Additionally, we mapped the sources to two other categories. First, the category \emph{whole process or approach} includes papers that cover multiple process phases. Second, the category \emph{other} includes various sub-categories in which multiple papers concentrated on a single theme. One paper can be mapped to multiple groups.

We found only two papers related to requirements engineering. In total, the software design category contains five papers and includes papers related to software acquisition. Papers related to programming and implementation, in general, are grouped under one category and included five papers. Papers related to integration, testing, and defect fixing are categorized under quality assurance and included 17 papers in total. For the process phases, papers related to maintenance are categorized with software evolution and release engineering and include three papers. Of all the sources categorized into software process phases, the highest number of papers is found in the quality assurance category. This could reflect that quality assurance is the last step before software releases, and thus, is performed just before the most critical deadlines.

Other sub-categories are derived from the qualitative coding with NVivo. We grouped papers related to whole processes or approaches into Agile and Scrum, project failure, and process improvement. In addition, papers related to temporal aspects of project management are in this category as subcategory cost estimation, cost models, simulation, and project escalation. There is a total of 29 papers in this category. The high number of studies related to cost estimation, cost models, simulation, and project escalation can be explained by the problems in these activities, as errors in cost estimation and scheduling cause time pressure. This is further explained in Section~\ref{sec:causes}.

Last, we put other groups under the group \emph{Other}. \emph{Detection of time pressure} includes papers that present or investigate ways of detecting hurry and arousal in software engineering. \emph{Group interaction} contains sources investigating interaction in software context. Sources not containing their own empirical evidence are grouped as \emph{Literature reviews and theoretical papers}. Papers from non-software engineering fields such as psychology, occupational health, and sociology as a fourth sub-category. We included these papers because they examine software engineering projects and offer valuable contributions to the understanding of time pressure in software engineering context. We also found three papers examining new product development and five papers that investigated group interaction in software engineering. We placed papers from studying factors related to the mind in the sub-category \emph{Individual psychological factors}.

\section{Empirical results}\label{sec:empiricalresults}

In this section, we present and summarize the empirical contributions of the papers we found in the literature. First, we review the identified causes of time pressure in Section~\ref{sec:causes}. Then, we present effects on individuals in Section~\ref{sec:individuals} and different software processes in Section~\ref{sec:process}. Finally, in section~\ref{sec:TCQS}, we review how time pressure affects the outcome of a software project by investigating the results through common project management measures of time, cost and quality. Each of these sections end with a summary helping a skimming reader.

\subsection{Causes of time pressure}\label{sec:causes}

\subsubsection{Effort estimation, scheduling, and management}

Several problems in effort estimation and scheduling that cause time pressure are mentioned in the literature. These problems include long schedules~\cite{jones2006social}, insufficient experience~\cite{verner2008factors}, insufficient historical data~\cite{smite2009cmmi}, schedule slips~\cite{van2013anatomy}, and change requests~\cite{lavallee2015good,jemielniak2009time}.

\citet{jones2006social} wrote about the reasons for software project failures. In his article, he listed the root causes of unrealistic schedule pressure as follows: ``1. Large software projects usually have long schedules of more than 36 months. 2. Project managers are not able to successfully defend conservative estimates. 3. Historical data from similar projects is not available. 4. Some kind of external business deadline affects the schedule.''.

In further work, \citet{ebert2009embedded} noted that projects with higher defect removal effectiveness tend to have shorter schedules, as testing is the part of development where delays typically happen. This observation is supported by Table~\ref{tab:phase} that shows a high number of papers in the quality assurance phase.  The authors elaborate further: ``Applications that enter testing with an excessive volume of defects cannot exit the testing phase because they don't work.''

 Incorrect estimates often lead to deviations from the initial project plan~\cite{basten2011current}. Another study showed that allowing schedule slippage increases time pressure if the final deadline is not adjusted~\cite{van2013anatomy}. Although the root cause of schedule pressure in \citet{van2013anatomy} was chronic under-staffing, the schedule slips increased the overall schedule pressure of this new product development project.

Reasons for effort estimation problems and subsequent failures to meet deadlines in Capability Maturity Model Integration (CMMI) level 5 organizations are given by \citet{smite2009cmmi}. In many cases, these problems stem from a lack of historical data for creating the estimates. Similarly, \citet{miranda2008protecting} suggested using probabilistic models to combat underestimation. In an older survey about using expert judgment as an estimation tool~\cite{hughes1996expert}, it was found that only few estimators use information about deadline pressure when producing estimates.

\citet{verner2008factors} investigation on the causes of failures showed that software projects failed because of multiple factors. However, three of the four most common factors involved time and were outlined as the ``delivery date impacted the development process,'' which was present in 93\% of failed projects, ``project was underestimated", which was present in 81\% of failed projects, and "staff not rewarded for working long hours", which was present in 73\% of failed projects. 

In a survey, partly by the same authors, time related reasons such as "project completed on time" are not seen as important for project success~\cite{procaccino2006defining}. In contradiction, a literature review summarizing project success factors~\cite{nasir2011critical} lists "Realistic schedule" as the third most common factor, and "Realistic budget" as the eight most common success factor on the reviewed literature.

Change requests, especially ones that are tied to internal dependencies of the developed software, have also been reported to increase time pressure~\cite{lavallee2015good,jemielniak2009time}. Similarly, requirements volatility has been mentioned as a cause for time pressure in a study conducted with surveys~\cite{ferreira2009understanding}. Overscoping, i.e. requiring more resources than available, is mentioned as a cause for time pressure by \citet{bjarnason2012you}. It has been noted by \citet{reichelt1999dynamics} that in complex projects with significant overruns, budgets were consumed for original work and any rework resulted in overruns and time pressure. Projects with budgets in which resources remained after initial rework were tied to less severe overruns~\cite{reichelt1999dynamics}.

Similarly, an unexpected shortage of resources, for example, unplanned leaves, have been reported to increase time pressures in software testing~\cite{shah2014global}. In the same paper, postponement of deadlines was also mentioned as one cause of time pressure. A case study investigating the effect of rapid releases on software testing found that testing becomes more deadline oriented with rapid releases, as testing is performed closer to deadlines~\cite{mantyla2013rapid}.

In the interviews, poor planning and a lack of organization were also mentioned as a cause of time pressure in addition to redundant meetings taking time away from other work tasks~\cite{deak2016challenges}. In interviews conducted by \citet{blackburn1996time}, managers mentioned that by not giving developers enough time, they forced developers to reuse more code. Last, interruptions and the increased cognitive load of task switching during a constantly changing software project was mentioned by \citet{sawyer2002temporal} as a reason for increased pressure and declining performance. Scheduling does not generally take individuals and their long-term tasks into account, resulting in difficulties prioritizing and multitasking when the task is continuously changing. 

To \textbf{summarize}, evidence shows that poor effort estimates lead to time pressure in software engineering. Conversely, realistic schedule is seen as a critical success factor for projects. A lack of historical data causes poor estimates, as well as business motivations for earlier deadlines, but more in-depth analysis for the reasons of poor estimates is beyond the scope of this work. If the final deadline is not adjusted, changes in a project's internal schedule do not help in dealing with time pressure. Moving deadlines up also increases time pressure. Based on \citet{ebert2009embedded} and the numerous studies in Table~\ref{tab:phase}, it appears that time pressure is most common in quality assurance. The lack of a buffer for unexpected work (e.g. change requests) or unexpected resourcing changes (e.g. unplanned leaves) leads to time pressure. Finally, poor organization of work and interruptions cause time pressure in software projects.

\subsubsection{Company culture}

Many papers reported time pressure and long hours as part of company culture, instead of time pressure being created due to shortcomings in effort estimation, which can be technical. Prolonged or constant pressure can lead to an unsustainable pace of software development and crisis mentality in the company.

\citet{perlow1998boundary} reported on the use of demands by senior managers to junior managers and engineers, where those
who demonstrate prioritization of work over their lives outside work are rewarded. This is also demonstrated by pressuring employees not to take time off during "crisis time," which leads to canceled vacations.

A company culture of individual heroics, high-visibility work and valuing of commitment to the company over everything were seen as factors that affect company culture and time pressure in another study by \citet{perlow1999time}. Individual heroics refer to a cycle observed in the company where deadlines are confronted too late and met with a crisis mentality and the extra effort of individual workers. High-visibility work was seen as a way to advance in the company; managers prioritized this work as they considered it crucial for their advancement, and resulted in regular checkups by the managers.

\citet{perlow2001time} also studied the time usage of software engineers in three different sites (China, India, and Hungary), to see if different cultural settings and management styles influence work hours or if long work hours are inherent in software engineering work. Perlow discovered variations in the way work is scheduled and coordinated, as well as in
the flexibility of when and where software engineers can work. Furthermore, Perlow found that specialized roles and personal modes of coordination make working hours more strict, as developers need to work more overlapping hours. 

\citet{perlow2002speed} conducted an ethnographic study of an Internet start-up during a period of 19 months, during which the company grew from a group of four students to a \$125 million company to bankruptcy. Because of the context of Internet start-ups, the company adopted a culture of fast decision making. Initially, it helped the company grow, but eventually, this mentality trapped the company in a process where they believed they had to make continually faster decisions to survive. The managers decided to "light a fire under the company" to create a "state of emergency address" to stimulate people. It created a sense of urgency which had a positive influence on the speed of decisions. Faster decisions created faster growth, which itself implied a need for faster decisions. This is the opposite of intuition and theory as, after initial growth, there should have been a lower sense of urgency according to previous work \cite{campion1982control}. The authors justified this difference with the context in which the company evolved: the fast world of the Internet.

\citet{tapia2004power} conducted a study using qualitative social research methods on the role of myths in the IT workplace. Tapia noted that in the company, employees working in teams challenged each other and developed a "one-up-man-ship'' culture, where employees competed to see who could spend the most time for work. This turned into more frequent
80-hour workweeks, as free time was expected after the next deadline but most of the time was not realized.

In a study by \citet{jemielniak2009time}, a similar company culture was reported, where managers believed programmers showed their commitment to the company by remaining at the workplace for longer hours. Additionally, in part because programmers' schedules were flexible at the company, and because programmers were asked to estimate how time-consuming their tasks were going to be, some software engineers admitted that scheduling and estimating changed to guessing the wishes of managers and the customer.

An action research study by ~\citet{borg2014ciar} focused on company culture in an ICT company located in Malta. Borg noted that using long work hours as a benchmark for ideal workers led to time pressure and even to burnout. Borg also noted the unequal effects on different kinds of workers, with mothers having the most trouble committing to the extra time demanded.

In \textbf{summary}, qualitative studies report on company cultures which foster time pressure. Factors reported to create company culture of constant time pressure include demanding prioritization of work over private lives to advance ones career \cite{perlow1998boundary, jemielniak2009time, borg2014ciar}, focus on individual heroics over development process  \cite{perlow1999time}, specialized roles and personal modes of coordination \cite{perlow2001time}, prioritizing a sense of urgency without a period of rest and refocus \cite{perlow2002speed}.

\subsection{Effects on individuals}\label{sec:individuals}

Positive effects of time pressure can include an increase in motivation or teamwork. \citet{paul2012time} report, based on an experiment, that in the context of short-term online virtual groups, time pressure enhances motivation. Similarly, \citet{marques2010coordination} reported that pressure acts as a support that triggers teamwork.

Two studies~\cite{ramanujan2000experimental, mantyla2014time} reported support for a mediating role of knowledge to perceived time pressure by an individual, that is, knowledge increases efficiency and decreases the effort needed for task completion.

\citet{langer2014project} and \citet{maruping2015folding} studied the effect of managers on team performance. \citet{langer2014project} studied the relation of managers' practical intelligence and job performance, finding that software projects with schedule pressure benefited from a manager who scored high on practical intelligence. Practical intelligence is related to ``resolving unexpected and difficult situations that often cannot be resolved using established processes and frameworks.'' Similarly, \citet{maruping2015folding} showed that ``managers can intervene to reorient team members' efforts toward effective task management through scheduling of interim milestones, synchronization of tasks, and restructuring of priorities,'', increasing team performance.

In the job demands-resources model, stress is assumed to be the result of an imbalance between demands and resources, for example, the demand for the tasks needed for the next deadline and the limited time resources before it. Hence many effects of stress can be linked to a lack of time. \citet{sanjram2013task} showed that programmers with time constraints experienced more significant workload as measured by the NASA-TLX assessment tool~\cite{hart2006nasa}. Furthermore, the group with time constraints failed more often when working on a separate task simulating multitasking. \citet{fehrenbacher2014behavioural} found that time pressure increases feelings of uneasiness and willingness to postpone decisions and decreases individuals' confidence.

Taking shortcuts has been linked to time pressure in software engineering. \citet{sojer2014understanding} reported that perceived severity of time pressure affects individuals' attitude toward unethical reuse of code, meaning developers who feel severe time pressure are more likely to have a more positive view on copying code unethically from the Internet. Similarly, on a survey by \citet{turley1995competencies}, shortcut taking was identified as the second most negative software development competence.

Agile ways of working have been reported to mediate the effects of time pressure. \citet{mann2005case} reported decrease of overtime work and an increase in customer satisfaction with the introduction of Scrum. \citet{laanti2013agile} reported that sustainable pace in Agile and Kanban projects lead to better performance, and that employees who reported being empowered were able to deal better with stress. \citet{tuomivaara2017short} found that developers who followed agile development process more closely felt less job strain at the end of the study period.

Time pressure was mentioned as the second most frequent reason for unhappiness among software developers in a survey conducted by \citet{graziotin2017unhappiness}. Additionally, time pressure was the most frequent cause of unhappiness from factors external to the developer. The most common reason was ``being stuck in problem-solving".

\citet{fujigaki1996time} investigated software engineers and their well-being by using the self-reported depressive scale (SDS) and semi-structured interviews to gauge job and life events. The self-reported depressive scale was developed by \citeauthor{zung1965self} in \citeyear{zung1965self} and has been widely used. \citet{fujigaki1996time} observed that SDS scores rose with increased job events, one of which was time pressure caused by deadlines. This result links time pressure to depressive symptoms, while higher depressive symptoms in turn have been linked to clinical depression. In a later study, \citet{fujigaki1997longitudinal} investigated physiological metrics in relation to the work strain of information system engineers. In the study adrenaline increased before a deadline, at the start of a project, and during budget negotiations. Cortisol, which captures exhaustion, increased after constant busy states, such as after deadlines and/or after employees had gotten used to the job.

\citet{borg2014ciar} observed that time pressure and increased working hours lead to burnout for individuals. Borg also noted the difficulty in balancing work and life outside of work and different attitudes between genders. Young mother reported being tired all the time, when her tasks continued at home with cleaning, preparing meals and taking care of the children. In contrast, men in the company mentioned their children in relation to leisure time.

\citet{mantyla2016mining} detected higher arousal (i.e., activity level) with sentiment analysis on more severe issues reports in JIRA repositories. Additionally, as issues are resolved, arousal drops, offering possible ways to detect time pressure

In \textbf{summary}, it has been shown that while time pressure can have positive effects on software developers, like increased motivation, it has also negative effects. These, such as increased stress and unhappiness, can eventually lead to depression and burnout. However, these negative effects on individuals can be mediated in different ways. In particular, Agile methods such as Kanban and Scrum decrease overtime and allows developers to better deal with stress and job strain. In addition, individuals' knowledge and managers' skills can also ease the negative effects of time pressure on individuals.

\subsection{Effect on the software process} \label{sec:process}

Speed accuracy trade-off~\cite{bakker2007job} and the covariance of decision speed with decision accuracy~\cite{heitz2014speed} can be seen as the general theory related to time pressure and software engineering. However, conflicting with general theory, the reported effects of time pressure on decision quality in software engineering literature are mixed. Less organizational change, communication, and knowledge transfer have been reported with time pressure. Additionally, time pressure has been reported to be an obstacle for software process improvement (SPI) and user involvement in the design process.

\subsubsection{Effect on software design} \label{sec:design}

\citet{rosenberg2017rapid} discuss schedule compression in resilient agile context at length, and consider the use of parallel development as the the most powerful practice. In the authors view, the amount of schedule which can be compressed is a trade-off between the amount of feedback and planning. The lowest cost is somewhere in the middle, i.e. having both some planning and feedback from the customer.

In a survey by \citet{tang2006survey}, the respondents considered the lack of time or budget to be the most common reason for not documenting design rationale. Similarly, \citet{rahmandad2009dynamics} report that under time pressure people tend to take shortcuts on documentation and requirements development. In a survey of the Chinese software industry, by \citet{yang2008survey} showed that one barrier to using software estimation cost models is schedule pressure. In the survey, the respondents proposed this answer; it was not a predefined answer option.

\subsubsection{Effect on communication and coordination} \label{sec:communication}

Time pressure also affects the communication and coordination within or outside the organization. Scientific evidence from software engineering suggests that with time pressure, there is more willingness to report bad news, less knowledge transfer, less communication between testers, and less organizational change.

Based on an experiment \citet{park2008overcoming} reported that individuals under time pressure were more willing to report bad news, as well as more likely to perceive themselves having personal responsibility to do so, which were in line with previous studies on the subject \citet{waller2001effect}.  However, less communication between developers and tester have been reported under time pressure ~\cite{shah2014global}. ~\cite{karhu2009investigating} found a negative relationship between knowledge transfer and schedule pressure, meaning those projects under study which reported knowledge transfer had less success staying in schedule and vice versa.

According to \citet{Staudenmayer2002} temporal shifts can be used as coordination mechanisms enabling organizational change. Temporal shifts consist of variations in five dimensions of how people experience time: a sense of time pressure, sense of ability to allocate time to different activities, perceived tension among competing task demands, the time horizon considered, and sense of found time. An introduction of buffer time during a project allowed for better review and reassessment of the project and subsequently for more organizational change.

\subsubsection{Effect on decision making}

The 19-month ethnographic study conducted by \citet{perlow2002speed} showed that although fast decision making was initially beneficial for the company's growth, it later led to bad decisions. With a growing artificial sense of emergency, decisions had to be made faster, and the decision-making board ignored valid objections because making a fast decision had become more important than making the right decision.

The effect of time pressure on decision making in Agile software development was pondered at length by \citet{riordan2013time}. In their conceptual framework, decision quality was explained with three temporal parameters: time pressure, polychronicity (i.e., unexpected or sporadic order of tasks), and iterative decision making.

\citet{lohan2014investigation} investigated decision quality under time pressure. Better decision quality was achieved when time pressure was perceived to be stimulating, enjoyable, or satisfying. However, based on results when time pressure is perceived to be annoying, discouraging, and upsetting, there does not appear to be an effect on decision quality.

\subsubsection{Effect in software process simulation models}\label{sec:models}

During our literature search, we included papers examining process simulation models and their schedule compression effects. Table~\ref{tab:models} presents the assumptions on the effects of time pressure underlying the models.

Two early studies~\cite{jeffery1987time, kitchenham1992empirical} compared empirical data with established cost estimation models taking schedule compression or extension into account, namely COCOMO I~\cite{boehm1981software} and \citeauthor{putnam1978general}'s~\cite{putnam1978general}. \citet{jeffery1987time} observed that, depending on the case, effort can either increase or decrease among 47 projects. Similarly, \citet{kitchenham1992empirical} noted two different schools of thought at the time. One where compressing or extending the schedule increases effort (COCOMO I), and one where compression increases effort and extension decreases it~\cite{putnam1978general}. Because of varying results, Kitchenham reports that all underlying assumptions are likely invalid.

\citet{smith2001empirical} added four task assignment factors to COCOMO I~\cite{boehm1981software} and produced more accurate estimations on a single project. One of them, \emph{intensity}, is defined as "the ratio between the number of active time units and the total number of time units during the development span". In the study, development effort has a negative relationship with intensity. This means that when development is focused on a single module, the overall effort on it decreases.

Later, the effects of different schedule compression levels on the estimates of COCOMO II~\cite{boehm2000cost} have been compared to real schedule compression ratios~\cite{yang2005effect, hussain2008effect}. \citet{yang2005effect} also present an overview of other cost estimation models and their schedule compression approaches (PRICE-S~\citet{Price1998}, SEER-SEM~\citet{fischman2005inside} and SLIM~\citet{panlilio1992software}). In COCOMO II compression increases the total effort needed for project completion. Based on newer empirical data, \citet{hussain2008effect} found an increased effort in highly compressed schedules.

Based on interviews, \citet{rahmandad2009dynamics} created a simulation model of the dynamics in concurrent software development. Developers under time pressure work harder, but also start to omit requirements, code reviews, unit testing, and documentation. This deteriorates quality and increases maintenance effort. In addition, organizational ability to produce software was reduced due to time pressure because not enough time was invested in improving the development tools or processes.

\citet{lin1997software} introduce a simulation model based on literature reviews, interviews and expert reviews. If the initial schedule estimation is compressed, it results in an overall increased effort. Otherwise an extended schedule corresponds to slightly lessened effort. \citet{hu1998software} produced a cost model based on a theory called Minimum Software Cost Model (MSCM),
where the overall cost in man-months, gets higher with longer schedules, other things being equal.

\citet{pfahl2001integrated} created multiple models by conducting interviews and participating in review meetings. In one of them, increased schedule pressure leads to defect injection, and faster work progress. Overall, the effect of time pressure can result in earlier completion of the project, but also in delays when errors are introduced because of schedule pressure.

An Agile process simulation model was created based on interviews and an extensive literature review by \citet{cao2010modeling}. Decreases in available time should result in adjustments in the scope or schedule, otherwise schedule pressure increases. Interviewees also stated that code refactoring was largely ignored and unit testing was reduced under schedule pressure, while pair-programming was said to perhaps alleviate the corner-cutting effects of pressure, as paired developers are usually more disciplined. No link between schedule pressure and development speed was suggested.

Another software process simulation model was developed by Abdel-Hamid and was discussed in multiple articles~\cite{abdel1990investigating, abdel1989lessons, abdel1990elusive}. The model is based on interviews and software managers' review of the resulting model. The main model proposed that schedule pressure leads to process losses and increases the error rate. However, the model does not have a link between schedule pressure and development rate, meaning it does not support the claim that schedule pressure improves development speed.

\citet{houston2001stochastic} simulated six risk factors of software development and produced a model called Software Project Actualized Risk Simulator (SPARS). The model assumed that the effects of excessive schedule pressure are fluctuating productivity, exhaustion, higher error creation, morale change, and weaker reviews. Based on previous models, \citet{ruiz2001simplified} present a reduced dynamic model (RDM) where schedule pressure increases errors and productivity.

\citet{van2010get} validated a software process simulation model in a new product development (NPD) software project, according to which schedule pressure increases errors, overwork, task rejection, and delays. The authors simulate an actual project which suffered from schedule pressure, and show that with a later due date, the overall effort would have been decreased.

\begin{table}[t]
\caption{Empirically derived assumptions of cost and simulation models on schedule compression}
\label{tab:models}
\begin{tabular}{p{10em}|p{7em}|l|l|l}
 & \multicolumn{2}{c|}{Compressed schedule} & \multicolumn{2}{c}{Eased schedule} \\
\hline
Model & Effort  & Quality & Effort & Quality \\
\hline
Putnam model and SLIM \cite{putnam1978general,panlilio1992software} & Increase & & Decrease & \\
\hline 
COCOMO I~\cite{boehm1981software} & Increase & & Increase & \\
\hline
PRICE-S~\citet{Price1998}& Increase & & Increase & \\
\hline
SEER-SEM~\citet{fischman2005inside}& Increase & & Increase & \\
\hline
System Dynamics Model \cite{abdel1990investigating} & Increase & Decrease & No Effect & \\
\hline
SEPS~\cite{lin1997software} & Increase & & Decrease & \\ \hline
MSCM~\cite{hu1998software} & Decrease & & Increase & \\
\hline
COCOMO II~\cite{boehm2000cost} & Increase & & No Effect & \\
\hline
SPARS~\cite{houston2001stochastic} & Both (usually increase) & Decrease & & \\
\hline
\citet{pfahl2001integrated} & Both & & & \\
\hline
RDM~\cite{ruiz2001simplified} & Decrease & Decrease & & \\
\hline
\citet{rahmandad2009dynamics} & Increase &  Decrease & & \\
\hline
\citet{cao2010modeling} & & Decrease & & \\
\hline
\citet{van2010get} & & Decrease & Decrease \\
\end{tabular}
\end{table}


In \textbf{summary}, most simulation and cost models report an increase in effort when the project's schedule is compressed, see Table \ref{tab:models}. While schedule compression can make developers work faster in the short term, it usually result in them omitting or avoiding reviewing, testing or documentating their code. This leads to an overall decrease in quality, more introduced errors and bugs, and eventually an increase in maintenance effort.

\mael{What to say about eased schedule? That the effects on effort are more mixed in the different studies?}

\subsubsection{Effect on software process improvement}

Several sources mentioned time pressure to be an essential obstacle in software process improvement (SPI)~\cite{paulish1994case, baddoo2000implementing, baddoo2003motivators}. However, practitioner surveys differed on the importance of time pressure as a barrier, with the more recent paper ranking time pressure lower in importance~\cite{niazi2005maturity, niazi2009software}.

\citet{paulish1994case} recommend using SPI techniques to improve the process to meet future deadlines without emergencies. Similarly, \citet{baddoo2000implementing} cite the lack of time as the biggest obstacle to software process improvement by participants from entry-level positions while strategic- and operational-level managers did not see the same importance for time pressure. \citet{baddoo2003motivators} report, based on a study of practitioner focus groups, that although both managers and developers reported the demotivating effect, the occurrence of time pressure as a demotivator was higher in focus groups composed of developers.

\citet{niazi2005maturity} identify time pressure as a barrier to implementing software process improvement (only 17\% of the time in interviews vs. 36\% in scientific literature, rank 2 vs. rank 5). In a later article~\cite{niazi2009software}, time pressure as a barrier to SPI is mentioned again. Resources should be explicitly allocated to SPI efforts to ensure adequate time to complete tasks.

\subsubsection{Effect on user involvement in the development process}

\citet{clegg1996software} present three case studies on software development in companies, one of which tried to include users in its software development process. An increase in time pressure decoupled the user and developer interaction, excluded user knowledge from the development process, and thus in the end, wasted resources and effort. However, afterwards the developers  stated that involving users in development could be improved with more realistic deadlines and better management.

\subsubsection{Effect on quality assurance}

\cite{baskerville2001internet} reported that time pressure lowers quality of the code during the initial product development and leading to rework and redesign during later product development. Similarly, high time pressure caused by unrealistic deadlines leads to minimal quality assurance~\cite{verner2008factors}. The quality assurance effort can be saved by using workarounds or compromises during implementation, by reducing the effort spent on documentation, by reallocating tasks to newly assigned developers, or by reducing the quality of the final product~\cite{basten2011current}. However, while the degree of time pressure seems to greatly reduce quality, \cite{hale2011evaluation} did not find an increased amount of defects on project with expedited schedules.

\subsubsection*{Effect on reviews} 

In a broad survey~\cite{ciolkowski2003software} about software reviews with 226 respondents from companies of different sizes and from different countries, time pressure was cited the most often (75\% of the time) as an obstacle to software review.

\citet{Staudenmayer2002} studied change in the development cycle in a big software company. The company introduced \emph{buffer times}. After several weeks of regular development, a buffer period of unallocated time is added. During the buffer time period, coding activity is suspended, but the tasks to be completed are not specified. Buffer time allows teams to reflect on the past and present, allowing the developers to switch from a development mode to a mode of reflection, awareness, and analysis. Developers can cope better with new or altered requirements caused by unexpected events, changes in customer needs, or other problems or ideas discovered during development. Teams in which buffer times were introduced kept to the schedule better and met release dates.

\subsubsection*{Effect on pair programming}

A study of pair programming of 31 developers from 4 companies~\cite{plonka2012disengagement} showed that time pressure (e.g., near the release date) leads developers to avoid pair programming and work individually to increase productivity.

\subsubsection*{Effect on testing}

In an experiment by \citet{mantyla2014time}, a group under time pressure found fewer defects, but the difference was not statistically significant. Overall time pressure was associated with higher efficiency (more defects found per unit of time). However, \citet{deak2016challenges} report a negative impact of time pressure on product quality, as well as its presence in Agile context when the project is behind schedule, as in the waterfall model.

\subsection{Effect on outcome - time-cost-quality scope}\label{sec:TCQS}

The so-called project management triangle dating back to the 1950s suggests that the outcome of any project can be explained by four constraints: quality, schedule, scope, and cost~\cite{atkinson1999project}. Time pressure in a project can be understood as a situation in which the project members realize that time, which can be either schedule (schedule pressure) or cost (time budget pressure), is running out. The project members try to avoid scheduling slippage or cost extension with various strategies, but typically by working faster. Studies have addressed quality, schedule, scope, and cost concerning time pressure. We divide the papers based on the data they use. We begin with studies with industrial project management data in Section~\ref{sec:industrialdata}, followed by experiments, surveys, and case studies in Sections~\ref{sec:experiments} through~\ref{sec:casestudy}. The results are summarized in table \ref{tab:descriptives}. 

For papers with quantitative data, we report whether time pressure has a positive or negative effect and statistical significance. We do not set an arbitrary threshold for significance reporting, often p = 0.05, as we consider this to be misleading. For example, if many sources all had p = 0.06, this would provide solid support for the impact of time pressure. Thus, omitting information based on an arbitrary threshold would not show this evidence.

\subsubsection{Industrial project management data} \label{sec:industrialdata}

Investigations of Capability Maturity Model (CMM) level 5 projects concerning effort, cycle time, and quality showed that schedule pressure decreases effort (p = 0.065) and cycle time (p = 0.15)~\cite{agrawal2007software}. In both cases, time pressure was the second most statistically significant predictor of the nine studies after project size. Schedule pressure was also linked to a decrease in quality but with a low p value (p = 0.4).   

\citet{hale2011evaluation} studied among other things, whether software projects with expedited schedules had more defects than projects with non-expedited in a Capability Maturity Model Integration (CMMI) level 3 company. However this hypothesis concerning schedule pressure was not supported (p = 0.400). \citet{mukhopadhyay1992software} investigated 58 software projects in the process control manufacturing domain and found that schedule pressure decreased software project effort with high statistical significance (p = 0.0001), while other statistically significant predictors were project size and programmer speed.

\citet{langer2014project} investigated the practical intelligence of project managers. The evidence showed that difficult projects achieved better client satisfaction and cost performance than standard projects because of the practical intelligence of the project manager. Concerning direct effects, the study reported that schedule pressure, surprisingly, increases cost (p $<$ 0.01) and reduces client satisfaction (p $<$ 0.01) even when they are controlled for project size.

\citet{nan2009impact} investigated the inverted U-shaped relationship (the initial pressure improves, but after a certain point, the pressure decreases performance) of budget and schedule pressure of 66 projects. They used regression models to predict the cycle time and development effort using budget and schedule pressure while controlling for the process maturity, size, complexity, and quality of the projects. For budget pressure, they found a U-shaped relationship. Furthermore, the linear terms of schedule pressure were negative, meaning an increase in schedule pressure reduced the cycle time and development effort. From the paper appendix, we found that quality had a negative correlation with budget pressure (-0.51), but a positive correlation with schedule pressure (0.51).

In earlier work, \citet{nan2003impact} conducted a similar investigation in a large company (\$1 billion/year). They found that time pressure (schedule) had a U-shaped relation with cycle time or effort. They also found that pressure had a non-statistically significant relationship with quality. However, this earlier work omitted many details, such as sample size and did not show the statistical values, making the results less trustworthy.

\citet{cataldo2010sources} reports that time pressure measured as concurrent execution of tasks was the most important source of errors (p $<$ 0.01) in distributed software development projects. In the regression model, the expected number of defects increased by 47.1\% when the value measuring time pressure changed from the minimum to maximum value. Similarly, \citet{cataldo2011factors} report that in global feature-oriented software development, time was the most significant factor when feature integration failed (p $<$ 0.01), as well as associated with a lower likelihood of failures in that feature. In the regression model based on a project with 1.5 million lines of code and 1,195 features, an additional week corresponds to a lower likelihood of 0.8\% of integration failure.

\subsubsection{Experiments}\label{sec:experiments}

A controlled experiment in requirements review and test case development~\citet{mantyla2014time} showed that time pressure reduced effectiveness in defect detection (p = 0.342) and had no impact on test case quality (p = 0.922). Efficiency, effectiveness divided by time, improved defect detection (p = 0.002) and test case quality (p $<$ 0.001). This result supports the idea that effectiveness (or quality of the work) will decrease, but efficiency will increase due to less time being used.  

An experiment on time pressure in manual testing by \citet{mantyla2013more} also showed lower effectiveness (p $<$ 0.001) and higher efficiency (p $<$ 0.001) under time pressure. The paper reports no t-test so we computed the p-values from the raw data. The researchers also concluded that combining several time-pressured testers would have been beneficial because ''we can either find roughly the same amount of defects with 59\% less effort, or we can use the same effort to find 71\% more defects." The drawback of using several independent time-pressured testers would have been the extra work in duplicate defect filtering. 

In another experiment, \citet{fehrenbacher2014behavioural} performed an experiment about time pressure in software acquisition. The study shows that under time pressure individuals worked faster but felt less confident in their decisions and were keener to postpone it. Gaze duration was reduced under time pressure as the individuals try to work faster (p = 0.02). Under time pressure, the work strategy is focused on higher-level topics, while on the other hand, without time pressure, more effort is spent looking at the details (p = 0.013). The best search of information for the software acquisition task occurred under time pressure and with requirements for explicitly written reasoning about the acquisition choice. Hence according to results, to achieve the best performance regarding effectiveness and time spent, time pressure and a quality control requirement should be used.

\citet{lee2012effect} showed in an experiment that project escalation, that is, willingness to continue a troubled software project, is less likely to happen if there is high time pressure in the project. This willingness to stop a problematic project was generally seen as positive by the authors as project escalation can waste valuable resources in "a failing course of action".

An experiment by \citet{ramanujan2000experimental} investigated the time used in short software maintenance tasks. For documented programs time pressure had no effect, but for programs with no documentation time pressure reduced maintenance task time by 20\%. The authors also found that the performance of participants with low and high knowledge increased (p = 0.0399) under time pressure. However, the impact was larger for low-knowledge participants (16\%) than for high-knowledge participants (6\%). The results offer further support that time pressure reduces effort. However, the finding that low-knowledge participants are affected more by time pressure is contrary to previous research.

Another controlled experiment by \citet{topi2005effects} of database query development tasks showed unexpectedly that time pressure didn't make the subjects work faster and reduced their efficiency. However, the p-values were high (p = 0.51 simple task, p = 0.82 complex task). Effectiveness was reduced under time pressure as well (p $<$ 0.001 simple task, p= 0.1281 complex task). The authors provide the following explanation: ''The subjects did not have good mechanisms for accelerating their work. Thus, this seems to indicate that with this task type, just reducing the available time does not improve productivity." On the other hand, time pressure reduced the number of total correct database queries.

\subsubsection{Surveys}

Investigation of software project teams reports a statistically significant inverted U-shaped relationship between team process and time pressure as the quadratic term of time pressure \citet{maruping2015folding}. The authors did not provide separate measures for product quality, or the effort used, but combined them in a team performance metric. However, temporal leadership statistically significantly removed the U-shaped relationship so that with strong temporal leadership, only positive effects of time pressure on team process exist. In the paper, temporal leadership was defined as ''the structuring, coordination, and management of task pacing in teamwork."

\citet{lohan2014investigation} investigated the effect of group cohesion, perceived challenge, and hindrance time pressure on decision-making quality of information system development. Challenge time pressure perceived as stimulating, enjoyable, and satisfying was thought to have a positive effect on quality. However, perceived hindrance time pressure did not have a negative effect.

In a survey, \citet{nugroho2007survey} found that meeting a deadline was considered by the respondents to be the smallest factor driving deviations between the design and code with 27\% of responses stating that the deadline never caused deviations. Of the chosen factors, meeting a deadline was the only factor that did not directly mention design quality (other factors being impractical design, incomplete design, and design not satisfying the requirements).

A survey conducted by \citet{verner2008factors} on software project failures found that too aggressive delivery dates caused time pressure, which then caused the omission of QA practices and led to project failures in six out of the eight projects studied. The authors proposed that projects should be kept short and manageable to prevent failures from too aggressive delivery dates. This practice sounds like Agile with small iterations. Another survey by \citet{ferreira2009understanding} showed that requirements volatility causes time pressure which increases errors in generating requirements.

\subsubsection{Case studies} \label{sec:casestudy} 

A qualitative case study by \citet{lavallee2015good} in a telecom company provided a concrete example of how taking shortcuts reduces overall quality. The researchers found that budget pressure prevented the implementation of a proper company-wide solution to a technical problem and resulted in a cheap patch solution that was repeated by at least 12 development teams. Each team was protecting their budget and decided to take a shortcut solution. A case study in another telecom company showed that time pressure was selected as the root cause for 40\% of the defects~\cite{leszak2000case}. A more detailed investigation showed that for algorithmic defects, the share of time pressure was as high as 70\%, while for functionality defect type it was only 17\%. The authors elaborated that "functionality defect refers to missing or wrong functionality (w.r.t. requirements) and algorithm defect refers to an inadequate (efficiency) or wrong (correctness) algorithmic realization."

\subsubsection{Summary}\label{sec:summary}

We summarize quantitative empirical evidence from analysing software companies and software engineering experiments that performed statistical tests regarding whether time pressure improves development efficiency and whether time pressure reduces development quality. For this, we consider only papers that measured actual outcomes. Thus, we omit the results of questionnaire surveys. Development efficiency means that software development is faster in terms of the cycle time, or effort, or both. In experiments, development efficiency means that effectiveness per time unit is faster, \emph{e.g.}, more defects per hour are found. Quality reduction in company cases was often measured by the number of defects or customer satisfaction. In experiments, quality reduction, becomes effectiveness, \emph{e.g.}, fewer defects are found in reviews, or less correct database queries are made. 

Table~\ref{tab:descriptives} shows the results. The \emph{+} sign means that the paper found an impact in the direction predicted in the column headings, that is, improved efficiency or reduced quality. The \emph{-} sign means the opposite, and \emph{U} sign means that some pressure results in the predicted impact, but too much pressure causes the opposite effect. This inverted U-shape effect comes from the Yerkes-Dodson law which states that initial pressure improves performance while pressure increasing above a certain point reduces the performance. After the sign, we report the statistical significance from the paper. The statistical significance can originate from various statistical tests, such as regression models, correlations, or t-tests. 

Seven papers support improvement in development efficiency due to time pressure, two papers report inverted U-shaped results, and two papers report a decreased efficiency. One of the two papers offering the counter-evidence had a strong statistical significance~\cite{langer2014project} but offered no explanation. Therefore we contacted the authors for further details but received no response. In the other paper, the statistical significance was much lower: p = 0.5 and 0.8. The paper also offered an explanation: in the case of database development tasks, the subjects were unable to go faster. Overall, we conclude from quantitative analysis of company data and experiments, that small to medium size time pressure is beneficial for development efficiency (9 paper for and 2 papers against). This is partly in contradiction to Section ~\ref{sec:models}, where majority of models assume the total amount of effort to increase with schedule compression.

Nine papers support a reduction in quality due to time pressure. Three papers report the opposite, that time pressure increased quality, and two papers provide no data on quality reduction. If we examine the three papers offering the counter-evidence, the statistical significance was very low on two papers p = 0.922~\cite{mantyla2014time} and p = 0.957~\cite{fehrenbacher2014behavioural}, while in the remaining paper~\cite{nan2009impact} with high statistical significance (p = 0.00) quality improvement computation was shown with a correlation only with industrial data. A regression model or partial correlation controlling for confounding factors would be a more robust alternative. We conclude from qualitative analysis of company data and experiments, that time pressure reduces quality in software engineering, but we suspect this is because less time is available or used. Reduced quality due to time pressure is further supported unanimously by the cost and simulation models in Section~\ref{sec:models}.

\begin{sidewaystable}[]
\centering
\caption{Summary on the effects of time pressure on efficiency and quality. Sign (+ , - or U) expresses the direction of relationship, i.e. whether time pressure improves (+) or decreases (-) efficiency, or whether they have non-linear relationship U-shaped (U).}
\label{tab:descriptives}
\begin{adjustbox}{width=\textwidth}
\begin{tabular}{l|l|l||l|l|l}
\textbf{Paper} & \textbf{N}   & \textbf{Data}                                     & \textbf{Pressure} & \textbf{Efficiency}                          & \textbf{Quality} \\
 &    &    & \textbf{Condition} & \textbf{Improvement}                          & \textbf{Reduction}\\
\hline
\cite{agrawal2007software}     & 31  & company projects                         & schedule           & +, p=0.065 (effort)                             & +, p=0.4 (defect count)                     \\
      &     &                                          &                    & +, p=0.15 (cycle time)                          &                                      \\
\cite{hale2011evaluation}     & 991  & company projects                         & schedule           &     NA                         & -, p=0.4 (increased amount of defects)                     \\
\cite{mukhopadhyay1992software}    & 58  & company projects                         & schedule           & +, p=0.0001*** (effort)                            & NA                                          \\
      &     &                                          &                    &                                                 &                                             \\
\cite{langer2014project}    & 530 & company projects                         & schedule           & -, p\textless{}0.01** (cost)                      & +, p\textless{}0.01** (client satisfaction)   \\
\cite{nan2009impact}     & 66  & company projects                         & budget             & U,  p=0.05*(cycle time)                            & -, p=0.00 (defect / size) correlation only  \\
      &     &                                          &                    & U, p=0.01** (effort)                                &                                             \\
\cite{nan2009impact}     & 66  & company projects                         & schedule           & +, p=0.01** (cycle time)                            & +, p=0.00 (defects / size) correlation only \\
      &     &                                          &                    & +, p=0.07 (effort)                                &                                             \\
\cite{nan2003impact}    & NA  & company projects                         & schedule           & U,  p\textless 0.01** (cycle time)               & ?, p = not significant                      \\
      &     &                                          &                    & U,  p\textless 0.05 (effort)                   &                                             \\
\cite{cataldo2010sources}          & 209    &  company projects    & concurrent tasks                    & NA                                                 &   +, p\textless 0.01** (defects)            \\
\cite{cataldo2011factors}              &  1195   & features in a project   &   time spent            &        NA             &   +, p\textless 0.01** (defects)             \\     
\cite{mantyla2014time}    & 97  & student experiment in & reward             & +, p\textless{}0.001*** (test case   & - , p=0.922 (test case score)               \\ 
      &     &      test case development                                    &                    &           score  / time)                                   &                                             \\
\cite{mantyla2014time}    & 97  & student experiment in    & reward             & +, p=0.002**               & +, p=0.342 (defects found)                  \\ 
&     &      requirement review                                    &                    &           (defects found / time)                              &                                             \\
\cite{mantyla2013more}$\dagger$   & 130 & student experiment in        & time-restriction   & +, p\textless{}2.2e-16***         & +, p=2.96e-09*** defects found)                \\ 
&     &      manual testing                                    &                    &                  (defects / time)                               &                                             \\
\cite{fehrenbacher2014behavioural}    & 106 & student experiment in  & time-restriction   & + p=0.020**                      & +, p=0.013** (number of fixations)            \\
&     &      software acquisition                                    &                    &                (fixation duration)                                 &          +, p=0.344 (number values)                                    \\
      &     &                                          &                    &                                                 &      -, p=0.957 (number of labels)            \\
\cite{ramanujan2000experimental}    & 100 & experiment software     & reward             & +, p=0.0399             & NA                                          \\
&     &      maintenance task                      &                    &             (time used, 2x2 Anova)                 &                                             \\
\cite{topi2005effects}$\dagger\dagger$  & 60  & experiment db query                               & time restriction   & -, p= 0.5138 (correct  & + p=0.0001*** (correct, simple task)     \\
&     &     &  & per minute, simple)  &  \\
      &     &                                          &                    & -, p= 0.8243 (correct & + p=0.1281 (correct, complex task) \\
      &     &     &  & per minute, complex)  &  \\
   \multicolumn{6}{l}{*p-values $<$ 0.05 highlighted with *, $<$ 0.01 with ** and $<$ 0.001 with ***} \\
   \multicolumn{6}{l}{$\dagger$ p-values missing from original paper computed from raw for this paper} \\
 \multicolumn{6}{l}{$\dagger\dagger$ we performed t-test from reported group means and standard deviations between low and high time pressure groups} \\
\end{tabular}
\end{adjustbox}
\end{sidewaystable}

\section{Threats to validity}\label{sec:threats}

The first threat to the validity of the findings is the search strings we used to query search engines for the literature. Before starting the literature review, we familiarized ourselves with the topic and iteratively improved the search strings. In total, we used synonyms and different ways of spelling, but it is possible some sources could have been missed with the otherwise inconsistent terminology used in the literature. However, as we did not run into other terms in the snowballing and analysis phase, we believe we have covered at least the terms most frequently used in the literature.

Several databases and search engines can be used to search scientific literature. Due to the limited amount of resources, we decided to use those that had the most extensive coverage, namely Scopus and Google Scholar. From these, partial automatic data retrieval is possible. However, it is possible that more sources of information could have been found by searching other academic databases, as data in a single one is incomplete and can contain errors such as missing abstracts. Indeed, we are aware of missing abstracts on some conference proceedings related to information systems. Another limiting factor in our study was using only first 100 search results for Google Scholar searches, which were meant to complement the primary searches made with Scopus. Indeed, more sources could possibly be found by increasing the number of search results examined for Google Scholar. As our resources are limited, and the application of the selection criteria more laborious than usual with reading at least the abstract, the boundary has to be set somewhere. We have tried to combat these issues with conducting backwards and forwards snowballing.

Although we applied the selection criteria specified in Section~\ref{sec:criteria} when we identified the relevant literature, we may have missed relevant papers. This is more the case in studies where time pressure was not the main topic investigated but constituted some of the empirical evidence. There is an inherent trade-off between the effort spent and the number of details that can be examined in the papers while applying the selection criterion.

The inexperience of the first author on performing systematic reviews can be mentioned as a threat to validity. However, other authors had previous experience on conducting systematic literature reviews and they provided guidance. First author had gained experience on conducting the previous work~\cite{kuutila2017reviewing}. Similarly, reviewer bias in study selection could be an issue. We tried to solve this as best as we could by marking even remotely borderline cases up for discussion as explained in Section ~\ref{sec:methodology}.


We could could have missed some details in the qualitative coding phase with NVivo due to errors in this stage. We are also aware of some more recent papers on the topic that were published during the analysis of the collected literature~\cite{salman2018effect,kuutila2018using}. However, adding the most recent papers to the analysis would be a never-ending circle. Last, we want to mention publication bias as a threat to these findings~\cite{sutton2000empirical}. It can be formulated that results, where no links between time pressure and investigated processes or approaches are found, might have a smaller chance of being published.
\section{Conclusions}\label{sec:conclusions}

We conclude the paper by first highlighting the contributions of this work. Next, we provide practical takeaways, and we outline directions for future work. 

\subsection{Contributions}
In this article, we perform the largest literature review related to time pressure in software engineering. Our main contributions are as follows.

\begin{itemize}
\item In Section~\ref{sec:definitions} we provide the definitions of time pressure as used in software engineering literature. They can be roughly be divided into two categories. The first one is based on the Yerkes-Dodson law, which states that the amount of time pressure affects performance in an inverted U-shaped form. Initial increases in pressure improve performance, although only up to a certain point, after which further increases in time pressure decrease performance. The second category of definitions is based on the challenge-hindrance framework, which states that positive (challenge) time pressure improves performance, while negative (hindrance) decreases performance. 

\item In Section~\ref{sec:metrics}, we provide a list of papers containing metrics and operationalizations, used in previous literature to measure and operationalize time pressure in software engineering.

\item In Section~\ref{sec:phase}, we map the selected papers to different stages of the software development process and approaches (see Table~\ref{tab:phase}). The main topics of papers related to time pressure were found to be either cost estimation or quality assurance.

\item In Section~\ref{sec:causes}, we summarize the reported causes of time pressure: problems in effort estimation, scheduling, commercial pressures, management styles, and social settings.

\item We review the effects of time pressure on individuals and software process outcomes. We summarize the corresponding quantitative results in Table~\ref{tab:descriptives}. The predominant effect of time pressure on outcome is that it reduces quality while increasing efficiency. However, many of the cost and process simulation models in Table~\ref{tab:models} support the increase in overall effort for a software project with compressed schedule.
\end{itemize}

\subsection{Practical takeaways}

We have performed many primary studies ourselves on performance under time pressure and arousal detection (e.g., ~\cite{mantyla2013more, kuutila2018using}). After spending considerable effort on systematically familiarizing ourselves with related literature, we want to conclude this paper by providing a practitioner-oriented summary with key takeaways.

Time pressure is common in the software industry, and it can be caused by commercial pressure, company culture, or errors in effort estimation (see Section~\ref{sec:causes}). Time pressure can have both positive and negative outcomes. On the positive side, it increases efficiency in the short term: the sense of urgency that time pressure creates provides focus on the basic product requirements. On the negative side, the lack of time reduces the quality of the outcome, leads to tunnel vision, and limits opportunities for improving the software product and process. See Section~\ref{sec:TCQS} and Table~\ref{tab:descriptives} for details.

The question thus becomes: ``can a software project achieve the best of both worlds: increased efficiency and urgency while avoiding reduced quality?" The answer is yes and no. No, in the sense that the best of both worlds cannot be achieved simultaneously. With heavy time pressure, it is difficult to find the time to make important improvements to the product during the software development process. However, it is possible for a project to have periods of time pressure and periods of buffer and reflection time, in which the former provides efficiency while the latter ensures a high quality of the product and process. This is a balance that skillful software engineers and managers should aim to achieve.

The amount and type of time pressure also play a role (see Section~\ref{sec:definitions}). Small to moderate time pressure brings out positive effects, while very high time pressure provides no additional benefits. If the pressure is experienced as positive, it leads to a more positive outcome than if the time pressure is experienced as negative. Positive time pressure can be achieved when a team feels that timely delivery is essential. Conversely, multiple conflicting goals, such as having to deliver a high-reliability product with minimal effort, can generate a negative time pressure. Typically, negative time pressure is more intense and has the trait of impossibility attached to it. Software engineers and managers should be mindful how the software development team feels about the time pressure they are dealing with. 

Saying that time pressure improves efficiency and reduces quality in software engineering is a generalization. Important task dependent variations are likely to occur. Although empirical evidence of task-dependent effects of time pressure in software engineering is limited, we found partial evidence for two variations. First, it appears that time pressure most often occurs during software quality assurance and testing (see Section~\ref{sec:phase}). This happens because testing, in particular, is the last phase that precedes software release: therefore, all the schedule slips of earlier phases are felt during software testing. Another possible cause is having a lower quality product because of time pressure, which ends up needing more testing. Second, the effects of time pressure vary according to the type of tasks. We found evidence that tasks with a high algorithmic nature have fewer efficiency improvements and suffer more from reduced quality than other types of tasks under time pressure~\cite{topi2005effects,leszak2000case}. Software engineers and managers should be particularly mindful in ensuring that software testing is not under too much negative time pressure and that tasks that are highly algorithmic in nature are under minimal pressure.

When it comes to different software process models, there is some empirical evidence and theoretical reasoning why Agile software development and time pressure are a good fit. We believe there are three reasons for this. First, in Agile software development, iterations are small, meaning that there is a constant low time pressure to meet the next deadline, although there is no massive final deadline~\cite{tuomivaara2017short, harris2009agile}. The sustainable pace of Agile is also good at cutting down extreme time pressure. Psychological experiments have shown that aggressive intermediate deadlines can improve outcome quality and individuals' time management~\cite{ariely2002procrastination}. We assume the same is true for software development teams. Second, in Agile software development, quality assurance and testing go hand-in-hand with development and thus, avoid the intense time pressure that haunts software testing in traditional phases. However, Agile is not a silver bullet, as lack of testing and re-factoring have been reported in Agile projects as well~\cite{deak2016challenges,cao2010modeling}. Third, team empowerment, which is higher in Agile than in traditional projects, can block the negative stress caused by time pressure~\cite{laanti2013agile}. This effect can be linked to the well-established occupational theory based on the job demands-control model~\cite{bakker2007job,karasek1979job}. This model proposes that adverse effects of time pressure (and other stressors) can be reduced when an employee has high independence and decision latitude in the job. This is precisely the case for Agile teams with high empowerment.

In more traditional development, processes can suffer from the effects of one final deadline, especially if there are no intermediate deadlines. Focusing on one final deadline can make the workload uneven, which has been linked to integration failures~\cite{cataldo2011factors}. Similarly, time budget pressure, combined with each team optimizing their own project, has been reported to lead to multiple teams developing a cheap patch instead of a company-wide solution~\cite{lavallee2015good}. 

\subsection{Takeaways for researchers}

There seems to be a contrast in results on productivity under time pressure between empirical studies on project performance and studies creating cost and process models, as it can be seen in Tables~\ref{tab:models} and~\ref{tab:descriptives}. The majority of cost and process models assume that overall effort increases with compressed schedules, whereas most empirical studies report improved efficiency under time pressure. There are multiple possible explanations for this. For example, in some of the models, the increased effort in compressed schedules is the result of an increase in error rate; thus, increased maintenance is needed to complete the project. Moreover, the time scale is also important: improved efficiency is more likely to occur in the short term, whereas negative effects become pronounced in the long term. Future studies could help establish guidelines for these trade-offs. 

Because of the U-shaped relationship between performance and arousal based on the Yerkes-Dodson law, quantifying the amount of time pressure on an individual would help investigate its effects at the project level data. The multiple assumptions and effects of schedule compression, as noted in Table~\ref{tab:models}, are can be partially explained by the difference in contexts in which software projects have been developed. However, the effects of time pressure on individuals might also play a role. Indeed, as it can be seen in Table ~\ref{tab:descriptives}, quantitative studies focused on experiments either on individuals, or on groups at the project level, but not on both at the same time. From the developers' point of view, it should matter if the system is developed with less re-usable code or with more overtime hours. This is further indicated by recent studies using the challenge-hindrance framework~\cite{lepine2005meta}. Taking into account these effects in future quantitative studies would provide better context and reasoning for conflicting results.

As noted in Section~\ref{sec:causes}, prior literature has identified company culture as a cause of time pressure. In these types of situations, hurry and time pressure are either prolonged or constant~\cite{perlow1999time, jemielniak2009time}. The negative effects of time pressure, such as increased stress, burnout, and depression, are reported to be products of exhaustion and job strain. Hence, the negative effects of time pressure might not become apparent in a single software project and especially in a lab experiment, if time pressure is otherwise at manageable levels or even rare in the company. We believe that company culture, together with varying time scales, might explain the discrepant assumptions and effects of schedule pressure and compression, noted in Tables ~\ref{tab:models} and \ref{tab:descriptives}.

Many recent studies have tried to detect time pressure with various techniques, including sentiment analysis~\cite{mantyla2017bootstrapping}, repository mining~\cite{kuutila2018using}, and physiological measurements~\cite{kolakowska2013emotion, tuomivaara2017short}. However, while promising results have been acquired, none of them have been able to reliably detect time pressure within a single project. Hence, if possible, future work should aim to accomplish this through combining different data sources. It could eventually provide project managers with up-to-date information on the state of a project, as felt by the individual developers.  
\section{Acknowledgements}\label{sec:acknowledgements}
This work has been supported by Academy of Finland grant 298020. The first author has been supported by Kaute-foundation.
\section{References}
\bibliographystyle{elsarticle-harv}
\bibliography{biblio}

\end{document}